%% file: main.tex
\documentclass[sigconf]{acmart}
\input{sty/usepackage_config.tex}

\AtBeginDocument{%
  \providecommand\BibTeX{{%
    \normalfont B\kern-0.5em{\scshape i\kern-0.25em b}\kern-0.8em\TeX}}}

\copyrightyear{2021} 
\acmYear{2021} 
\setcopyright{acmcopyright}
\acmConference[IMC '21]{ACM Internet Measurement Conference}{November 2--4, 2021}{Virtual Event, USA}
\acmBooktitle{ACM Internet Measurement Conference (IMC '21), November 2--4, 2021, Virtual Event, USA}
\acmPrice{15.00}
\acmDOI{10.1145/3487552.3487814}
\acmISBN{978-1-4503-9129-0/21/11}

\begin{document}

\providecommand{\sysname}{{\sc TopoShot}\xspace}

\title{\sysname: Uncovering Ethereum's Network Topology Leveraging Replacement Transactions}
\renewcommand{\shortauthors}{Kai Li et al.}

\pagenumbering{gobble}



\author{Kai Li}
\email{kli111@syr.edu}
\affiliation{%
  \institution{Syracuse University}
  \city{Syracuse}
  \state{NY}
  \country{USA}
}

\renewcommand{\footnotemark}{{\ \Envelope}}
\author{Yuzhe Tang}
\authornote{\ \Envelope\ \xspace{}Yuzhe Tang is the corresponding author.}

\email{ytang100@syr.edu}
\affiliation{%
  \institution{Syracuse University}
  \city{Syracuse}
  \state{NY}
  \country{USA}
}

\author{Jiaqi Chen}
\email{jchen217@syr.edu}
\affiliation{%
  \institution{Syracuse University}
  \city{Syracuse}
  \state{NY}
  \country{USA}
}

\author{Yibo Wang}
\email{ywang349@syr.edu}
\affiliation{%
  \institution{Syracuse University}
  \city{Syracuse}
  \state{NY}
  \country{USA}
}

\author{Xianghong Liu}
\email{xliu317@syr.edu}
\affiliation{%
  \institution{Syracuse University}
  \city{Syracuse}
  \state{NY}
  \country{USA}
}

\newcommand{\ignore}[1]{}
\input{text/abstract.tex}

\ccsdesc[500]{Networks~Peer-to-peer protocols}
\ccsdesc[500]{Networks~Peer-to-peer networks}

\keywords{Blockchain, Overlay networks, Network measurements, Ethereum transactions}

\settopmatter{printfolios=true}
\maketitle

\input{text/mainbody.tex}
\bibliographystyle{ACM-Reference-Format}
\bibliography{ads,lsm,bkc,odb,cacheattacks,sc,crypto,sgx,diffpriv,txtbk,distrkvs,vc,yuzhetang}


\clearpage
\appendix
\input{text/wellformatted_appendix.tex}

\end{document}

%% file: sty/usepackage_config.tex
\usepackage{amsfonts}

\usepackage{subfig}
\usepackage{wrapfig}
\usepackage{float}
\usepackage{multirow, graphicx}
\usepackage{algorithm}
\usepackage{algpseudocode} 
\usepackage{amsmath}

\usepackage{xspace}

\usepackage{url}

\usepackage{sty/widetext} 

\usepackage{stackrel}

\let\emptyset\varnothing

\ifdefined\TTSTYLETARGETreport
\usepackage[
    backend=bibtex,
    maxbibnames=100,
    firstinits=true,
    bibstyle=numeric,
    citestyle=numeric
]{biblatex}
\fi

\usepackage{pifont}
\newcommand{\cmark}{\ding{51}}%
\newcommand{\xmark}{\ding{55}}%

\input{sty/wheels-usepackage.tex}

\usepackage[misc]{ifsym} 
\usepackage{bbding} 

\usepackage{todonotes}
\usepackage{libertine}
\usepackage{balance}

%% file: sty/wheels-usepackage.tex
\usepackage{graphicx, multirow}

\usepackage{tabularx, multirow, rotating} 
\usepackage{subfig} 
\usepackage{algorithm} 
\usepackage{algpseudocode} 
\usepackage{ulem} 
\usepackage{color} 

\usepackage{listings} 
\lstdefinestyle{Oracle}{basicstyle=\ttfamily,
                        keywordstyle=\lstuppercase,
                        emphstyle=\itshape,
                        showstringspaces=true,
                        }
\makeatletter
\newcommand{\lstuppercase}{\uppercase\expandafter{\expandafter\lst@token
                           \expandafter{\the\lst@token}}}
\newcommand{\lstlowercase}{\lowercase\expandafter{\expandafter\lst@token
                           \expandafter{\the\lst@token}}}
\makeatother

\usepackage{tikz}

\newcommand{\ballnumber}[1]{\tikz[baseline=(myanchor.base)] \node[circle,fill=.,inner sep=1pt] (myanchor) {\color{-.}\bfseries\footnotesize #1};}

\newif\ifboldnumber

\algrenewcommand\alglinenumber[1]{%
  \footnotesize\ifboldnumber\bfseries\fi\global\boldnumberfalse#1:}


%% file: text/abstract.tex
\begin{abstract}
Ethereum relies on a peer-to-peer overlay network to propagate information. The knowledge of Ethereum network topology holds the key to understanding Ethereum's security, availability, and user anonymity. 
However, an Ethereum network's topology is stored in individual nodes' internal routing tables, measuring which poses challenges and remains an open research problem in the existing literature.

This paper presents \sysname, a new method uniquely repurposing Ethereum's transaction replacement/eviction policies for topology measurement. \sysname can be configured to support Geth, Parity and other major Ethereum clients. As validated on local nodes, \sysname achieves 100\% measurement precision and high recall ($88\%\sim{}97\%$). To efficiently measure the large Ethereum networks in the wild, we propose a non-trivial schedule to run pair-wise measurements in parallel.
To enable ethical measurement on Ethereum mainnet, we propose workload-adaptive configurations of \sysname to minimize the service interruption to target nodes/network.

We systematically measure a variety of Ethereum networks and obtain new knowledge including the full-network topology in major testnets (Ropsten, Rinkeby and Goerli) and critical sub-network topology in the mainnet. 
The results on testnets show interesting graph-theoretic properties, such as all testnets exhibit graph modularity significantly lower than random graphs, implying resilience to network partitions. The mainnet results show biased neighbor selection strategies adopted by critical Ethereum services such as mining pools and transaction relays, implying a degree of centralization in real Ethereum networks.
\end{abstract}

\ignore{
Ethereum relies on a peer-to-peer overlay network to propagate information. The knowledge of Ethereum network topology holds the key to understanding Ethereum's security, availability, and user anonymity. From a measurement perspective, an Ethereum network's topology is routing-table information hidden inside individual Ethereum nodes, measuring which poses challenges and remains an open research problem in the existing literature.

This work presents TopoShot, a new method uniquely repurposing Ethereum's transaction replacement/eviction policies for topology measurement. TopoShot can be configured to support Geth, Parity and other major Ethereum clients. As validated on local nodes, TopoShot achieves 100\% measurement precision and high recall ($88\%\sim{}97\%$). To efficiently measure the large Ethereum networks in the wild, we propose a non-trivial schedule to run pair-wise measurements in parallel.
To enable ethical measurement on Ethereum mainnet, we propose workload-adaptive configurations of TopoShot to minimize the service interruption to target nodes/network.

We systematically measure a variety of Ethereum networks and obtain new knowledge including the full-network topology in major testnets (Ropsten, Rinkeby and Goerli) and critical sub-network topology in the mainnet.
The results on testnets show interesting graph-theoretic properties, such as all testnets exhibit graph modularity significantly lower than random graphs, implying resilience to network partitions. The mainnet results show biased neighbor selection strategies adopted by critical Ethereum services such as mining pools and transaction relays, implying a degree of centralization in real Ethereum networks.
}

%% file: text/mainbody.tex
\newcommand{\tremark}[1]{\footnote{\textcolor{red}{(Ting's comment: #1)}}}
\newcommand{\xremark}[1]{\footnote{\textcolor{red}{(Xin's comment: #1)}}}
\newcommand{\kl}[1]{\footnote{\textcolor{blue}{(Kai: #1)}}}
\newcommand{\yz}[1]{\footnote{\textcolor{red}{(Yuzhe: #1)}}}
\newcommand{\tangSide}[1]{\todo[caption={},color=cyan!20!]{{\scriptsize #1}}}

\providecommand{\infura}{{SrvR1}\xspace}
\providecommand{\blockdaemon}{{SrvR2}\xspace}
\providecommand{\etherscan}{{SrvR3}\xspace}

\providecommand{\sparkpool}{{SrvM1}\xspace}
\providecommand{\twominers}{{SrvM2}\xspace}
\providecommand{\icanminingru}{{SrvM3}\xspace}
\providecommand{\digipools}{{SrvM4}\xspace}

\providecommand{\cheapethpool}{{SrvM5}\xspace}
\providecommand{\ubiqpool}{{SrvM6}\xspace}

\definecolor{mygreen}{rgb}{0,0.6,0}
\lstset{ %
  backgroundcolor=\color{white},   
  basicstyle=\scriptsize\ttfamily,        
  breakatwhitespace=false,         
  breaklines=true,                 
  captionpos=b,                    
  commentstyle=\color{mygreen},    
  deletekeywords={...},            
  escapeinside={\%*}{*)},          
  extendedchars=true,              
  keepspaces=true,                 
  keywordstyle=\color{blue},       
  language=Java,                 
  morekeywords={*,...},            
  numbers=left,                    
  numbersep=5pt,                   
  numberstyle=\scriptsize\color{black}, 
  rulecolor=\color{black},         
  showspaces=false,                
  showstringspaces=false,          
  showtabs=false,                  
  stepnumber=1,                    
  stringstyle=\color{mymauve},     
  tabsize=2,                       
  title=\lstname,                  
  moredelim=[is][\bf]{*}{*},
}

\input{text/paper.tex}

\section*{Acknowledgments}
The authors thank the anonymous reviewers in ACM IMC'21, SIGCOMM'21, and SIGMETRICS'21.
The authors are partially supported by the National Science Foundation under Grants CNS1815814 and DGE2104532.

\clearpage

%% file: text/paper.tex
\section{Introduction}

\ignore{A blockchain system relies on an underlying peer-to-peer (P2P) network to propagate information including recent transactions and blocks. Knowing the topology of P2P network is essential to understanding blockchain's availability under network partitions, its security against denial of service attacks (e.g., eclipse attacks~\cite{DBLP:conf/uss/HeilmanKZG15} and denial of specific node services~\cite{DBLP:conf/ndss/LiCLT0L21,me:deter}), and its protection of user anonymity~\cite{DBLP:conf/ccs/BiryukovKP14,DBLP:conf/fc/KoshyKM14}. The value has motivated a line of measurement studies on the network topology of popular blockchains including Bitcoin~\cite{DBLP:conf/fc/Delgado-SeguraB19,DBLP:conf/fc/GrundmannNH18} and Monero~\cite{DBLP:conf/fc/CaoYDLV20}. 
}

A blockchain system relies on an underlying peer-to-peer (P2P) network to propagate information including recent transactions and blocks. The topology of the P2P network is foundational to the blockchain's availability under network partitions, its security against a variety of attacks (e.g., eclipsing targeted nodes~\cite{DBLP:conf/uss/HeilmanKZG15}, denial of specific node service~\cite{me:deter,DBLP:conf/ndss/LiCLT0L21}, and deanonymization of transaction senders~\cite{DBLP:conf/ccs/BiryukovKP14,DBLP:conf/fc/KoshyKM14}), and its performance (e.g., mining power utilization~\cite{DBLP:conf/fc/GencerBERS18} and the quality of RPC services~\cite{me:infura,me:etherscan,me:quiknode}). Details are in \S~\ref{sec:topo:significance}. This value has motivated a line of measurement studies on the network topology of popular blockchains including Bitcoin~\cite{DBLP:conf/fc/Delgado-SeguraB19,DBLP:conf/fc/GrundmannNH18} and Monero~\cite{DBLP:conf/fc/CaoYDLV20}.
However, although Ethereum is the second largest blockchain network (after Bitcoin) and the biggest smart-contract platform, measuring Ethereum's network topology remains an open research problem. The existing Ethereum measurement studies~\cite{DBLP:conf/imc/KimMMMMB18,kiffer2021under} focus on profiling individual peer nodes, but not the connections among them.

\begin{figure}
\centering
  \includegraphics[width=0.25\textwidth]{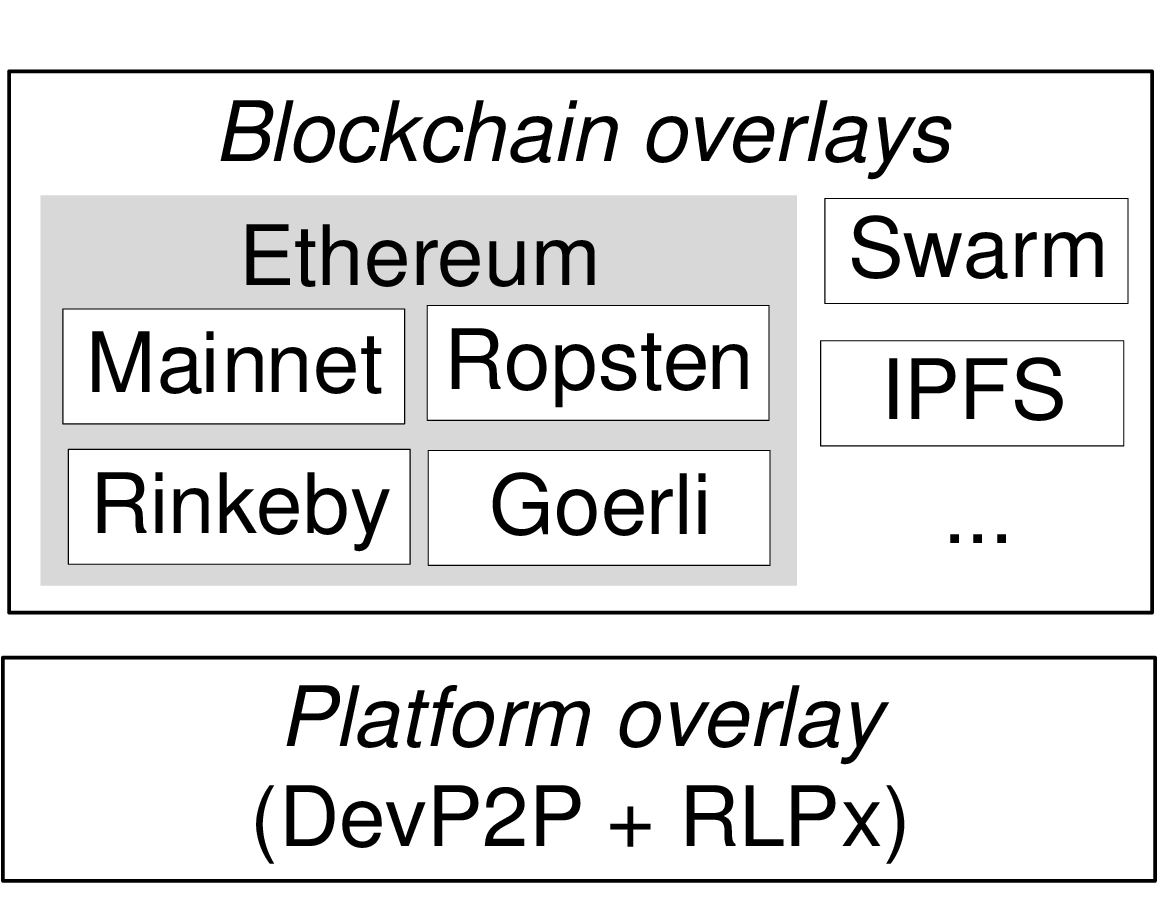}%
  \caption{P2P overlay networks on Ethereum. Shaded is the measurement target of this work.}
  \label{fig:ethereum:overlays}
\end{figure}

\vspace{3pt}\noindent\textbf{Research goals}:
Specifically, the operational Ethereum P2P network today runs tens of thousands nodes and host multiple overlays: 1) an underlying P2P overlay, called platform overlay, which forms a structured DHT network by following Kademlia's protocols~\cite{DBLP:conf/iptps/MaymounkovM02} for peer discovery (RLPx) and session establishment (DevP2P)~\cite{DBLP:conf/imc/KimMMMMB18}, and 2) a number of application-specific overlays~\cite{me:swarm,DBLP:conf/imc/KimMMMMB18}, among which the dominant ones are Ethereum blockchains for information propagation. In particular, the Ethereum P2P network hosts multiple blockchain overlays with different ``networkIDs'' including the mainnet and various testnets, such as Ropsten~\cite{me:ropsten}, Rinkeby~\cite{me:rinkeby} and Goerli~\cite{me:goerli}. This multi-layer view of Ethereum's P2P network is depicted in Figure~\ref{fig:ethereum:overlays}.
In the P2P network, each Ethereum node maintains ``peer'' connections at these two layers: 1) On the blockchain overlay, a node maintains a list of {\it active} neighbors through which local information is propagated. 2) On the platform overlay, a node stores the {\it inactive} neighbors in a DHT routing table, from which live nodes are promoted to active neighbors in the future.

This work aims at measuring the Ethereum P2P network's blockchain overlay and its {\it active} links.\footnote{In this paper, we use terms ``links'', ``connections'' and ``edges'', interchangeably.} 
In practice, it is the blockchain overlay's active links, instead of platform overlay's inactive ones, that capture the exact flow of information propagation and are more informative. For instance, a node running the Geth client~\cite{me:geth} (which is the most popular Ethereum client and is deployed on more than $80\%$ nodes in the mainnet~\cite{me:ethernodes}) maintains $272$ inactive neighbors and around $50$ active neighbors, by default. Knowing what these 50 neighbors are is helpful to understand the node's resilience to eclipse attacks (as information is propagated through the $50$ active neighbors, not the $272$ inactive ones, and an attacker only needs to disable the $50$ active neighbors to block information propagation). Also, knowing whether the 50 neighbors contain nodes from top mining pools, such as Sparkpool~\cite{me:sparkpool} (or popular transaction relay service, such as infura.io~\cite{me:infura}), is useful to estimate the timeliness and quality of the blocks (or transactions) received on the node, as well as understanding the centralization of the blockchain network.

Measuring Ethereum network's active links is an open research problem. In the existing literature, 
1) the related works that measure Ethereum networks focus on profiling individual nodes~\cite{DBLP:conf/imc/KimMMMMB18,kiffer2021under} or detecting inactive links~\cite{DBLP:conf/iscc/0003SWTZY19,DBLP:journals/corr/abs-2104-03044}, but not the active connections. 
Compared to the inactive links that are exposed in Ethereum peer discovery messages (i.e., RLPx's \texttt{FIND\_NODE} packets) and can be directly measured as in~\cite{DBLP:conf/iscc/0003SWTZY19,DBLP:journals/corr/abs-2104-03044}, active links are hidden information inside remote Ethereum nodes, directly measuring which without inference is impossible as we thoroughly examine Ethereum protocol's messages.
2) The other related works explore the topology of non-Ethereum blockchains including Bitcoin~\cite{DBLP:conf/fc/Delgado-SeguraB19,DBLP:conf/fc/GrundmannNH18} and Monero~\cite{DBLP:conf/fc/CaoYDLV20}. 
Their measurement approaches exploit features specific to Bitcoin/Monero and are inapplicable to Ethereum as will be detailed in \S~\ref{sec:rw}. Notably, TxProbe's approach~\cite{DBLP:conf/fc/Delgado-SeguraB19} to infer Bitcoin's topology cannot be applied to measuring Ethereum topology, as these two blockchains differ in transaction model (account-based versus UTXO-based) and propagation model (direct propagation versus announcement), as will be further explained in \S~\ref{sec:txprobe:applicable}.

\vspace{3pt}\noindent\textbf{Measurement methods}:
In this work, we propose \sysname to measure an Ethereum blockchain overlay by repurposing Ethereum's transaction replacement and eviction policies. 
Briefly, an Ethereum node buffers unconfirmed transactions (prior to mining) in a local data structure named \texttt{mempool}, where an unconfirmed transaction can be replaced or evicted by a subsequent transaction at a sufficiently higher Gas price.\footnote{The difference between transaction replacement and eviction is that a transaction $tx$ is replaced by another transaction from the same sender account with $tx$, and $tx$ is evicted by another transaction from a different sender from $tx$ when the mempool is full.}
Transaction replacement and eviction are standard Ethereum features, widely supported by Ethereum clients (including Geth~\cite{me:geth}, Parity~\cite{me:parity} and others~\cite{me:besu,me:aleth,me:nethermind}), and highly desirable by real-world applications. For instance, a common practice in blockchain-based decentralized applications is that a user having sent a transaction can posthumously speed up its inclusion into the blockchain by sending replacement transactions at higher price per computation unit (or the so-called Gas price).
Leveraging these features, \sysname runs a measurement node $M$ to detect the connection between two remote nodes $A$ and $B$. 
In \sysname, node $M$ propagates a high-priced transaction $tx_A$ on target node $A$, a low-priced transaction $tx_B$ to target node $B$, and a medium-priced transaction $tx_C$ propagated to all other nodes in the same network. 
It then observes $tx_A$'s presence on node $B$ and, if so,  draws the conclusion that node $A$ is actively connected to node $B$. 
To ensure the accurate measurement, when node $A$ is not linked to node $B$, measurement transaction $tx_A$ should not be propagated and do not reach node $B$ (the so-called ``isolation'' property~\cite{DBLP:conf/fc/Delgado-SeguraB19}). 
One of the key insights in this work is that Ethereum's transaction replacement policy can be repurposed to enforce isolation property for accurate link measurement. Intuitively, the isolation is ensured by the fact that \sysname's high-priced $tx_A$ can replace the low-priced $tx_B$ on node $B$ but not the medium-priced $tx_C$ on other nodes, through which $tx_A$ cannot be propagated to reach node $B$.

To set up the measurement as above, \sysname further leverages Ethereum's  support of transaction eviction and future transactions, that is, to evict an existing unconfirmed transaction on a node by incoming future transactions (the concept of future transaction in Ethereum is similar to orphan transactions in Bitcoin). {
Specifically, when using \sysname to measure the connectivity between Nodes $A$ and $B$, the measurement node $M$ first needs to connect to both nodes, propagates $tx_C$ to all nodes, then sends future transactions to evict $tx_C$ (with other existing transactions) on node $A$ and $B$ before sending $tx_A$ and $tx_B$ to node $A$ and $B$, respectively. This method can be applied to measuring the connectivity among all possible pairs of nodes by the standard approach of launching a ``supernode'' connecting to all other nodes in the network~\cite{DBLP:conf/imc/KimMMMMB18,kiffer2021under}.}

The basic \sysname achieves $100\%$ result precision but not $100\%$ recall, which can be attributed to non-default settings of target node. We further propose a pre-processing phase retrofittable with \sysname to profile the actual settings of target node and to improve the result recall, proactively. 

For large-scale measurement on real Ethereum networks, we propose a non-trivial method to parallelize multiple pair-wise measurements, reducing the rounds and overall time of measurement.

\vspace{3pt}\noindent\textbf{Measurement results}:
We systematically evaluate the validity of \sysname and conduct measurement studies on both testnets and the mainnet. The measurement results uncover, for the first time, the full network topology of Ethereum's major testnets (including Ropsten, Rinkeby and Goerli) and the inter- and intra-connections among the mainnet's mining-pools and transaction relay services. 

First, we validate the \sysname's correctness in terms of recall and precision. We set up a local node under control and a remote node in a testnet, and we use \sysname to measure the connection between the two nodes. By comparing against the ground-truth of node connection (via querying the local node's state), we confirm that \sysname achieves the perfect precision ($100\%$) and a high recall (up to $97\%$).

Second, we use \sysname to measure, for the first time, the network topology of major Ethereum testnets including Ropsten, Goerli and Rinkeby. We also analyze the captured graphs which reveal  a number of graph-theoretical properties including degree distribution, distances, assortativity, clustering and community structures. Our comparative analysis shows that the measured Ethereum networks have particularly lower modularity than classic random graphs~\cite{Erdos60onthe,newman2003structure,albert2002statistical}, implying a better resilience against attacks to partition the networks.

Third, we propose enhanced \sysname configurations to allow lightweight yet effective measurement on the mainnet without ethical concerns.
The \sysname enhancement minimizes the impacts on the target nodes being measured, and particularly ensures that the set of transactions included in the blockchain do not change under measurement.
Using the approach, we measure a critical substructure of Ethereum's mainnet overlay. The result reveals biased neighbor selection strategy commonly practiced by critical Ethereum services such as mining pools and transaction relays who prioritize to connect other critical nodes over average nodes.
We acknowledge the high cost of our method and avoid measuring the topology of entire mainnet network which would otherwise cost $60$ million USD at the Ether price as of May 2021.

\vspace{3pt}\noindent\textbf{Contributions}: This work makes the following contributions:

\vspace{2pt}\noindent$\bullet$\textit{
Novel methods}: We propose a novel method, named \sysname, to measure Ethereum network links and topology. \sysname takes a unique approach by exploiting Ethereum's handling of unconfirmed transactions (i.e., transaction replacement and eviction). \sysname is generic and supports all Ethereum clients (including Geth and Parity). \sysname is effective and achieves 100\% result precision and high recall ($88\%\sim{}97\%$).

\vspace{2pt}\noindent$\bullet$\textit{
Large-scale measurements}: We address the scalability and ethical challenges raised in measuring large-scale, real Ethereum networks. We propose to schedule pair-wise measurements in parallel for efficiency. We propose workload-adaptive mechanisms to configure \sysname for minimal service interruption on the target nodes/network.

\vspace{2pt}\noindent$\bullet$\textit{
New systematic results}: Without \sysname, an Ethereum network's topology remains hidden information inside blackbox Ethereum nodes, measuring which stays an open research problem. By systematically conducting measurements against a variety of Ethereum networks, we obtain a series of new knowledge on network topology and its graph-theoretic statistics, ranging from full-network topology in popular testnets (Ropsten, Rinkeby and Goerli) and critical sub-network topology in the mainnet.

{The source code of \sysname is on \url{https://github.com/syracuse-fullstacksecurity/Toposhot}.
}

\vspace{3pt}\noindent\textbf{Roadmap}: The paper is organized in the following order: \S~\ref{sec:prel} presents the preliminary knowledge. Motivation of this work is presented in \S~\ref{sec:topo:significance}. \S~\ref{sec:rw} surveys the related works and their (in)applicability to measuring Ethereum's topology. \S~\ref{sec:method} presents \sysname's measurement primitive, parallel schedule, as well as correctness analysis. \S~\ref{sec:results} presents the measurement results of Ethereum testnets and the mainnet. The ethical aspects of this work are discussed in \S~\ref{sec:discuss}, and the conclusion is in \S~\ref{sec:conclude}.

\section{Preliminary}
\label{sec:prel}

{\bf Ethereum transactions}: To begin with, we describe the transaction model used in Ethereum. An Ethereum transaction binds a sender account to a receiver account. Each transaction is associated with a {\it nonce}, which is a monotonically increasing counter per sender. 
An Ethereum transaction is associated with {\it Gas price}, that is, the amount of Ether the sender is willing to pay to the miner for each unit of computation carried out by the miner to validate the transaction.

{\bf Unconfirmed transaction buffer (\texttt{mempool})}:
Each Ethereum node stores unconfirmed transactions in a local data structure, named \texttt{mempool}. In a \texttt{mempool}, a transaction $tx$ is {\it pending}, if its nonce equals one plus the maximal nonce of the transactions of the same sender in the \texttt{mempool} (i.e., equal to $n+1$). Otherwise, if $tx$'s nonce is strictly larger than $n+1$, $tx$ is a {\it future transaction}.

When a transaction $tx$ propagated from other nodes arrives at a node $N$, node $N$ determines whether to {\it admit} $tx$ into its \texttt{mempool}. 
Admitting a transaction $tx$ may trigger two more \texttt{mempool} events: 1a) {\it eviction} of an existing transaction $tx'$ by $tx$ where $tx$ and $tx'$ are of different accounts or nonces, and 1b) {\it replacement} of an existing transaction $tx'$ by $tx$ where $tx$ and $tx'$ are of the same sender and nonce. 

{\bf Transaction propagation}:
When admitting a pending transaction to its \texttt{mempool}, an Ethereum node propagates the transaction to its active neighbors. If an incoming transaction is not admitted or the admitted transaction is a future transaction, it will not be propagated. 

Normally, an Ethereum node directly {\it pushes} a pending transaction to its neighbors. That is, it sends a message to its neighbors encoding the transactions it wants to propagate. It may be the case that the propagated transactions are already received on the neighbors. This is the default transaction propagation protocol supported widely in Geth, Parity and other clients.

Some Ethereum clients (e.g., Geth with version later than 1.9.11) support {\it announcements} as an optional transaction-propagation protocol.
It works in three messages: 1) The node announces its local pending transactions by their hash and propagates the hash to its neighbors. 2) Then within the next $5$ seconds, its neighbors will respond with requests if they want to receive the transaction. Within these $5$ seconds, the neighbors will not respond to other announcements of the same transaction. 3) The node propagates the transactions to all requesting neighbors.
While this is similar to Bitcoin's transaction announcement as exploited in TxProbe~\cite{DBLP:conf/fc/Delgado-SeguraB19}, there is an important distinction: Ethereum's transaction announcement has to {\it co-exist} with transaction pushing, and Ethereum's pushing can bypass the blocking of an announcement.


{\section{Motivation: Significance of Knowing Blockchain Topology}
\label{sec:topo:significance}

The motivation of this work is that a blockchain network's topology is foundational to the blockchain's security and performance. In this section, we present a non-exhaustive list of ``use cases'' of blockchain topology knowledge in the hope of justifying its significance. 

\subsection{Implication to Blockchain Security}
The knowledge of blockchain network topology is crucial to understanding its security against various attack vectors, with examples listed below.

{\bf Use case 1: Targeted eclipse attacks}.
In the network topology, if a blockchain node is found to be of a low degree (i.e., few neighbors), such a node is particularly vulnerable under a targeted eclipse attack~\cite{DBLP:conf/uss/HeilmanKZG15}.
That is, such an eclipse attacker can concentrate her attack payload to the few neighbors to disable the information propagation and to isolate the victim node from the rest of the network at low costs.

{\bf Use case 2: Single point of failure}. The blockchain's network topology may reveal the centralization of network connection, leading to a single point of failure. Specifically, there may be supernodes that connect to all other nodes, ``bridge'' nodes that control the connection to the backend of critical services, and topology-critical nodes removing which may lead to partitioned networks. Directing denial-of-service attacks onto these critical nodes, using attack vectors recently discovered~\cite{me:deter,DBLP:conf/ndss/LiCLT0L21}, can lead to consequences such as crippled blockchain services and the censorship of individual transactions.

{\bf Use case 3: Deanonymizing transaction senders}.
With the knowledge of the network topology, if nodes' neighbors are distinguishing (i.e., node $X$'s neighbors are distinct from another node $Y$'s neighbors), the neighbor set can be used to identify/fingerprint nodes and can be further used to facilitate the deanonymization of transaction senders. Specifically, in the deanonymization attack~\cite{DBLP:conf/ccs/BiryukovKP14}, a blockchain ``client'' node (i.e., a node behind the NAT) is identified by its ``server''-node neighbors (a server node is of public IPs, is not behind the NAT and accepts incoming connections). An attacker then simply monitors the transaction traffic on all server nodes in the blockchain (e.g., a Bitcoin network contains much fewer server nodes than the client nodes, thus lowering the attacker's costs). The attacker can link a transaction sender's blockchain address (her public key) to a client node's IP address, which can be further linked to a real-world identity, thus deanonymizing the blockchain address.  

\subsection{Implication to Blockchain Performance}
Blockchain network topology is essential to achieving its performance promises and matters to both miners and client users. 

{\bf Use case 4: Mining efficiency and mining pools' QoS (quality of service)}.
In a blockchain, the time to propagate a recently found block from its miner to the entire network is critically important: If it takes too long to propagate miner $X$'s block, her block may arrive after another miner $Y$'s block, leading to $X$'s block unable to be included in the blockchain and $X$'s loss of revenue. Thus, a blockchain's network topology that affects propagation delay can influence a miner node's revenue and mining-power utilization (i.e., how much mining power spent is useful and is reflected in the main chain's blocks). Thus, it is unfavorable to have a minor with limited connectivity and incur long propagation delays. 

For a client interested in joining a mining pool, she may want to access the knowledge of blockchain topology and make an informed decision to choose the mining pool with better connectivity and lower propagation delay to ensure high mining revenue.

{\bf Use case 5: RPC service's QoS}.
For a client sending transactions through RPC services (e.g., infura.io), she may want to choose a service with better connectivity so that her transaction can be relayed on a timely basis.

In summary, the knowledge of blockchain network topology is critical to understanding its security, performance, and decentralization. Given the high market capitalization of today's blockchains (e.g., $\$4106$ billion USD for Ethereum as of Sep. 2021~\cite{me:coinmarketcap:ethereum}), we believe measuring blockchains' network topology is valuable and worthy even it may cost as much as $60$ million USD, estimated in \S~\ref{sec:mainnet}. 
}

\section{Related Work}
\label{sec:rw}
In this section, we present the existing measurement studies on public blockchain networks. Existing works can be categorized into three classes: W1) Measuring blockchain nodes, W2) measuring blockchain inactive edges, and W3) measuring blockchain active edges.

\noindent
{\bf Measuring blockchain nodes (W1)}: Kim et al.~\cite{DBLP:conf/imc/KimMMMMB18} propose a passive method to characterize the Ethereum mainnet by launching a ``supernode'' to connect all reachable mainnet nodes and collecting messages exchanged. The result reveals node-wise characteristics including network size, node geo-distribution, clients' age and freshness, and others.

Neudecker et al. (2019)~\cite{Neudecker2019_1000091933} is a passive measurement study that last four years to characterize the behavior of individual Bitcoin peers and their operators. Their approach is by launching ``supernodes'' and passively collecting transaction traffic, a method similar to~\cite{DBLP:conf/imc/KimMMMMB18}. 

\noindent
{\bf Measuring blockchain inactive edges (W2)}: Ethereum's peer discovery protocol (RLPx) has a \texttt{FIND\_NODE} message through which a node can discover another node's current routing-table entries (inactive neighbors). Recent research works~\cite{DBLP:conf/iscc/0003SWTZY19,DBLP:journals/corr/abs-2104-03044} directly measure Ethereum's inactive links by sending  \texttt{FIND\_NODE} messages to all nodes in an Ethereum network. This method cannot distinguish a node's ($50$) active neighbors from its ($272$) inactive ones and does not reveal the exact topology information as \sysname does. 

Henningsen et al.~\cite{DBLP:conf/networking/HenningsenFR020} measure the Kademlia network topology in IPFS by sending crafted peer-discovery queries. Despite other findings, this work reveals IPFS's network combines a structured Kademlia DHT and an unstructured P2P overlay. 

\noindent
{\bf Inference of blockchains' active edges (W3)}: 
Coinscope~\cite{miller2015discovering} targets Bitcoin's network topology and infers the links by leveraging the expiration timestamps of Bitcoin's \texttt{ADDR} messages.

TxProbe~\cite{DBLP:conf/fc/Delgado-SeguraB19} infers Bitcoin's network topology by exploiting Bitcoin's support of orphan transactions and announcement-based transaction propagation. We will describe how TxProbe works, with more details in \S~\ref{sec:txprobe:applicable},  to understand its applicability to measuring Ethereum networks.

Grundmann et al.~\cite{DBLP:conf/fc/GrundmannNH18} present two Bitcoin-topology inference approaches among which the more accurate one exploits Bitcoin's behavior of dropping double-spending transactions.
Neudecker et al. (2016)~\cite{DBLP:conf/uic/NeudeckerAH16} conducts a timing analysis of Bitcoin transaction propagation to infer the network topology. Despite the optimization, both works are limited in terms of low accuracy. 

Daniel et al.~\cite{DBLP:conf/lcn/DanielRT19} propose to exploit block relay mechanisms to passively infer connections among mining nodes and their direct neighbors in the ZCash network. 

Cao et al.~\cite{DBLP:conf/fc/CaoYDLV20} measure Monero's P2P network topology by exploiting the timing of neighbors' liveness probes. Specifically, a Monero node maintains the liveness of its neighbors (the \texttt{last\_seen} label) by periodically discovering its hop-2 neighbors, probing their liveness by sending PING messages, and selectively promoting them to be hop-1 neighbors. Topology inference methods are proposed to exploit the timing difference of neighbor nodes' \texttt{last\_seen} labels. This method is specific to  Monero's liveness-check protocol.

\begin{table}[!htbp] 
\caption{Existing works on blockchain topology measurement and \sysname's distinction}
\label{tab:distinct}\centering{\small
\begin{tabularx}{0.425\textwidth}{ |X|l|l|l| }
\hline
Research work	&Blockchain	&Measurement target \\\hline
\cite{Neudecker2019_1000091933}	&Bitcoin	&Nodes\\\hline
TxProbe~\cite{DBLP:conf/fc/Delgado-SeguraB19}
\& others~\cite{miller2015discovering,DBLP:conf/fc/GrundmannNH18,DBLP:conf/uic/NeudeckerAH16}
	&Bitcoin	&Active edges (W3)\\\hline
\cite{DBLP:conf/fc/CaoYDLV20}	&Monero	& Active edges (W3)\\\hline
\cite{DBLP:conf/lcn/DanielRT19}	&ZCash	& Active edges (W3)\\\hline
\cite{DBLP:conf/imc/KimMMMMB18}	&Ethereum	& Nodes (W1)\\\hline
\cite{DBLP:conf/iscc/0003SWTZY19,DBLP:journals/corr/abs-2104-03044}	&Ethereum	& Inactive edges (W2)\\\hline
\cite{DBLP:conf/networking/HenningsenFR020}	&IPFS	& Inactive edges (W2)\\\hline
\sysname	&Ethereum	& Active edges (W3)\\\hline
\end{tabularx}
}
\end{table}

The existing blockchain measurement studies are summarized in Table~\ref{tab:distinct}. In general, existing techniques on W1 and W2 directly measure the target (as the target information of nodes and inactive edges is exposed in collected messages), while measuring active edges (W3) requires inference. Existing topology-inference techniques focus on non-Ethereum blockchains and exploit blockchain specific features (e.g., Monero's timing of liveness probes and Bitcoin's announcement-based propagation) that are absent in Ethereum.


\subsection{TxProbe's Applicability to Ethereum}
\label{sec:txprobe:applicable}

To understand how TxProbe works and its (in)applicability to measuring Ethereum network, we first describe the following measurement framework: Suppose a measurement node $M$ is to detect the connection between a pair of target nodes, say $A$ and $B$. Node $M$ can propagate a transaction $tx_A$ to node $A$ and observe $tx_A$'s presence on node $B$. If present, nodes $A$ and $B$ are actively linked. The success of this method depends on the so-called isolation property. That is, when node $A$ and $B$ are not actively linked, $tx_A$ should not be propagated to node $B$. In other words, there is no alternative routing path beside the direct link between $A$ and $B$ that $tx_A$ can take to reach node $B$. 

TxProbe~\cite{DBLP:conf/fc/Delgado-SeguraB19,DBLP:conf/fc/GrundmannNH18} materializes this framework to measure active links in Bitcoin and ensures the isolation property by Bitcoin's transaction announcement mechanism. Briefly, Bitcoin's transaction announcement works as follows: a Bitcoin node propagates a transaction to its neighbor by first sending a transaction announcement (i.e., a hash value) to the neighbor and, upon neighbor's acknowledgment, then sending the actual transaction. Bitcoin has a  policy that the neighbor node receiving an announcement will ignore the subsequent announcements of the same transaction from other nodes for $120$ seconds. TxProbe exploits this policy to ensure the isolation during the $120$-second period. This is done by having Node $M$ to announce $tx_A$ to all nodes besides $B$ so that these nodes will not relay $tx_A$ when Node $A$ starts to propagate $tx_A$ to $B$, ensuring the isolation property.

However, TxProbe's method is inapplicable to measuring Ethereum. Ethereum's transaction propagation only {\it partially} depends on announcement, that is, a transaction is announced to some neighbors and is directly sent to other neighbors without announcement. The existence of direct propagation, no matter how small portion it plays, negates the isolation property, as measurement transaction $tx_A$ can be propagated through the nodes taking direct propagation as the alternative path to reach node $B$, introducing false positives to the measurement results.

In addition, TxProbe relies on Bitcoin's UTXO model, which differs from Ethereum's account model. Directly applying TxProbe to Ethereum risks incorrect measurement, as analyzed in Appendix~\ref{appdx:sec:txprobe:applicable}.

\section{\sysname Measurement Methods}
\label{sec:method}

We first present our observation on real Ethereum clients' behavior in transaction replacement and eviction, which lays the foundation of \sysname measurement method (\S~\ref{sec:mempoolprofile}). We then describe the measurement primitive in \sysname that detects just one link between two nodes (\S~\ref{sec:measure:serial}). We will then describe how to use this primitive to measure a network of a large number of links (\S~\ref{sec:measure:par}).

\subsection{Profiling Ethereum Clients' Behavior}
\label{sec:mempoolprofile}

We first describe a parameterized model for \texttt{mempool} and then our measurement study that reveals the \texttt{mempool} parameters of real Ethereum clients.

{\bf \texttt{mempool} model}: 
Recall that transaction eviction (replacement) is a \texttt{mempool} process that takes as input the initial state of \texttt{mempool} and an incoming transaction $tx_1$ and produces as output the end state of the \texttt{mempool} and possibly, a transaction $tx_2$ that is of the same (different) sender with $tx_1$ and that is evicted (replaced) from the \texttt{mempool}. To formally describe the process, suppose the initial state is a full \texttt{mempool} consisting of $l$ pending transactions and $L-l$ future transactions, where $L$ is the capacity of the \texttt{mempool} (denoted in Table~\ref{tab:notate}). The incoming transaction $tx_1$ is a future transaction with Gas price higher than any transactions currently in the \texttt{mempool}. There are $u$ transactions currently in the \texttt{mempool} that are of the same sender with $tx_1$.

When there is another transaction $tx_2$ in the \texttt{mempool} that has the same sender and nonce with $tx_1$, admitting $tx_1$ to the \texttt{mempool} triggers the replacement of $tx_2$. The generic transaction replacement strategy is that {\it \texttt{mempool} decides to replace $tx_2$ by $tx_1$, if $tx_1$'s Gas price is $1+R$ of $tx_2$'s Gas price}.

Otherwise (i.e., when there is no transaction of the same sender and nonce with $tx_1$), admitting $tx_1$ may trigger transaction eviction. 
For transaction eviction, the situation of interest to us is the eviction victim $tx_2$ being a pending transaction. Under this situation, the transaction eviction strategy generally follows the template: {\it \texttt{mempool} decides to evict a pending transaction $tx_2$ by $tx_1$, if 1) $tx_1$'s Gas price is higher than $tx_2$'s Gas price, and 2) there are more than $P$ pending transactions existing in the \texttt{mempool}, and 3) there are fewer than $U$ existing transactions of the same sender with $tx_1$.}
The three \texttt{mempool} parameters, namely $R$, $U$ and $P$, and their meanings are presented in Table~\ref{tab:notate}.

\begin{table}[!htbp] 
\caption{Notations}
\label{tab:notate}\centering{\footnotesize
\begin{tabularx}{0.5\textwidth}{ |l|X| }
 \hline
 Notation & Meaning \\ \hline
$R$ & Minimal Gas price difference for an incoming transaction (tx) to replace an existing tx in \texttt{mempool} \\ \hline
$U$ & Max number of future txs sent from the same account that can be admitted to a node's \texttt{mempool} \\ \hline
$P$ & Minimal number of pending txs buffered in a node to allow eviction by future txs \\ \hline
$L$ & Maximal number of txs allowed to store in a \texttt{mempool} (\texttt{mempool} capacity) \\ \hline
\end{tabularx}
}
\end{table}

\begin{table}[!htbp] 
\caption{Profiling different Ethereum clients in terms of transaction eviction and replacement policies 
The second column refers to the percentage of mainnet nodes running a specific client~\cite{me:ethernodes}.}
\label{tab:measure:diffclients}\centering{\small
\begin{tabularx}{0.5\textwidth}{ |X|X|X|l|l|l| }
\hline
{Ethereum clients} & {Deployment (mainnet)} & Replacement behavior & \multicolumn{3}{l|}{Eviction behavior}
\\ \cline{3-6}
& & $R$  & $U$  & $P$ & $L$
\\ \hline
Geth            & $83.24\%$     & $10\%$        & $4096$        & $0$& $5120$\\ \hline
Parity          & $14.57\%$     & $12.5\%$      & $81$          & $2000$& $8192$\\ \hline
Nethermind      & $1.53\%$      & $0\%$         & $17$          & $0$   & $2048$ \\ \hline
Besu            & $0.52\%$      & $10\%$        & $\infty$      & $0$   & $4096$ \\ \hline
Aleth           & $0\%$         & $0\%$         & $1$           & $0$   & $2048$ \\ \hline
\end{tabularx}
}
\end{table}

\label{sec:unittest}
\noindent{\bf \texttt{mempool} tests}: 
The measurement is set up with 1) a measurement node $M$ running the test and 2) a target node $T$ running the Ethereum client to be measured.
For each test, node $T$'s initial state of \texttt{mempool} contains $l$ future transactions and $L-l$ pending transactions. 

We design the first set of tests to trigger transaction replacement and measure $R$.
Specifically, $tx_1$ has an identical sender and nonce with an existing transaction $tx_2$ in \texttt{mempool}. In each unit test, Node $M$ sends $tx_1$ of a certain Gas price to node $T$, and observes if node $T$ replaces $tx_2$ by $tx_1$. We run a series of unit tests with varying $tx_1$'s Gas prices, in order to observe the minimal Gas price that triggers the replacement, from which we calculate and report $R$.

We design the second set of tests to trigger transaction eviction and measure $U$ and $P$.
Specifically, the \texttt{mempool} contains $L-l$ future transactions and $l$ pending transactions, among which there are $u$ transactions sent from the same account with future transaction $tx_1$.
As before, in each unit test, node $M$ sends to node $T$ $tx_1$ at a Gas price higher than any transactions in node $T$'s \texttt{mempool}.
We run a series of unit tests with varying $l$ and $u$. We observe the maximal value of $u$ that triggers a successful eviction by $tx_1$ and report such value by $U$. We observe the minimal value of $l$ that triggers a successful eviction by $tx_1$ and report such value by $P$.

\noindent{\bf Test results on Ethereum clients}: 
We conduct the tests on two local nodes: We first set up a local measurement node $M$ running tests on an instrumented Geth client and a local target node $T$. The statically instrumented Geth client allows node $M$ to bypass local checks and to propagate future transactions to node $T$.
We run the two sets of tests against target node $T$ running five different Ethereum clients: Geth (Go), OpenEthereum/Parity (Rust), Nethermind (.net), Besu (Java) and Aleth (C++).
Here, we discard the Python client (i.e., Trinity) as the incomplete implementation. The distribution of mainnet nodes running the five Ethereum clients is presented in the second column of Table~\ref{tab:measure:diffclients}, where Geth ($83\%$) and Parity ($15\%$) are the dominant clients on the mainnet.

The measurement results are reported in Table~\ref{tab:measure:diffclients}.
The \texttt{mempool} model and measurement results will guide the design of \sysname's method and the configuration of the measurement on different Ethereum clients.
Noteworthy here is that Aleth's and Nethermind's $R$ values are both zero ($0\%$), which renders our \sysname unable to work (as will be seen, it requires a non-zero $R$ to enforce the isolation property). Thus, \sysname currently does not work with Aleth and Nethermind clients. On the other hand, we deem a zero-value $R$ is a flawed setting that can be exploited to construct low-cost denial of service or flooding. For instance, an attacker can send multiple replacing transactions at almost the same Gas price, consuming network resources by propagating multiple transactions yet without paying additional Ether. We sent bug reports to Ethereum Foundation's bug bounty program~\cite{me:ethfoundation:bugbounty}, and further updates, if any, will be documented on a private link~\cite{me:toposhot:bugreport2}.

\begin{figure*}[!ht]
  \begin{center}
    \subfloat[Workflow: {Shaded are transactions in the nodes' \texttt{mempool}. Step 2 would evict $tx_C$ and add $tx_B$ on node $B$'s \texttt{mempool}, and Step 3 evicts $tx_C$ and adds $tx_A$ on node $A$'s \texttt{mempool}.}]{
  \includegraphics[width=0.45\textwidth]{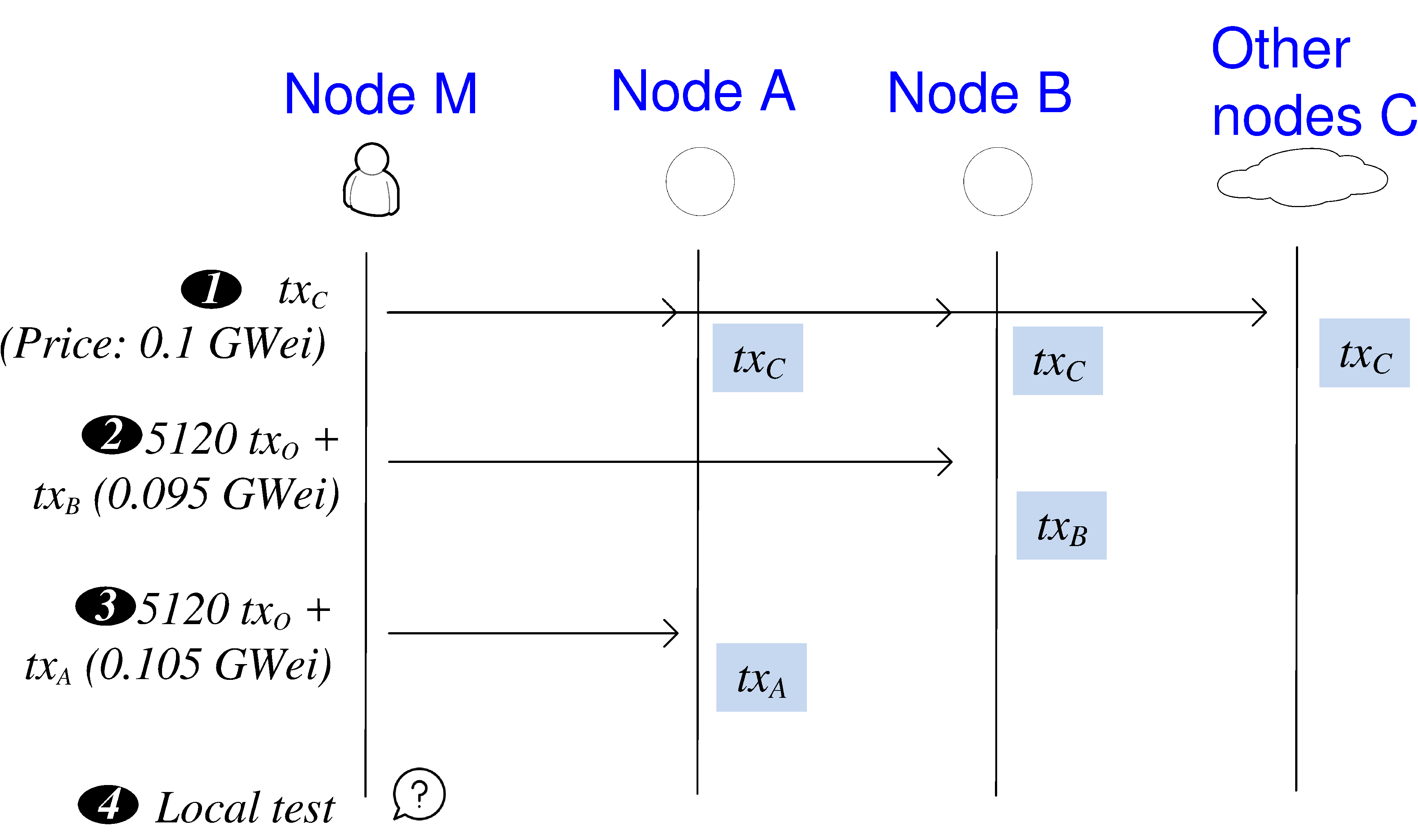}%
  \label{fig:workflow}
    }%
\hspace{0.4in}
    \subfloat[Snapshot right before Step 4]{
  \includegraphics[width=0.3\textwidth]{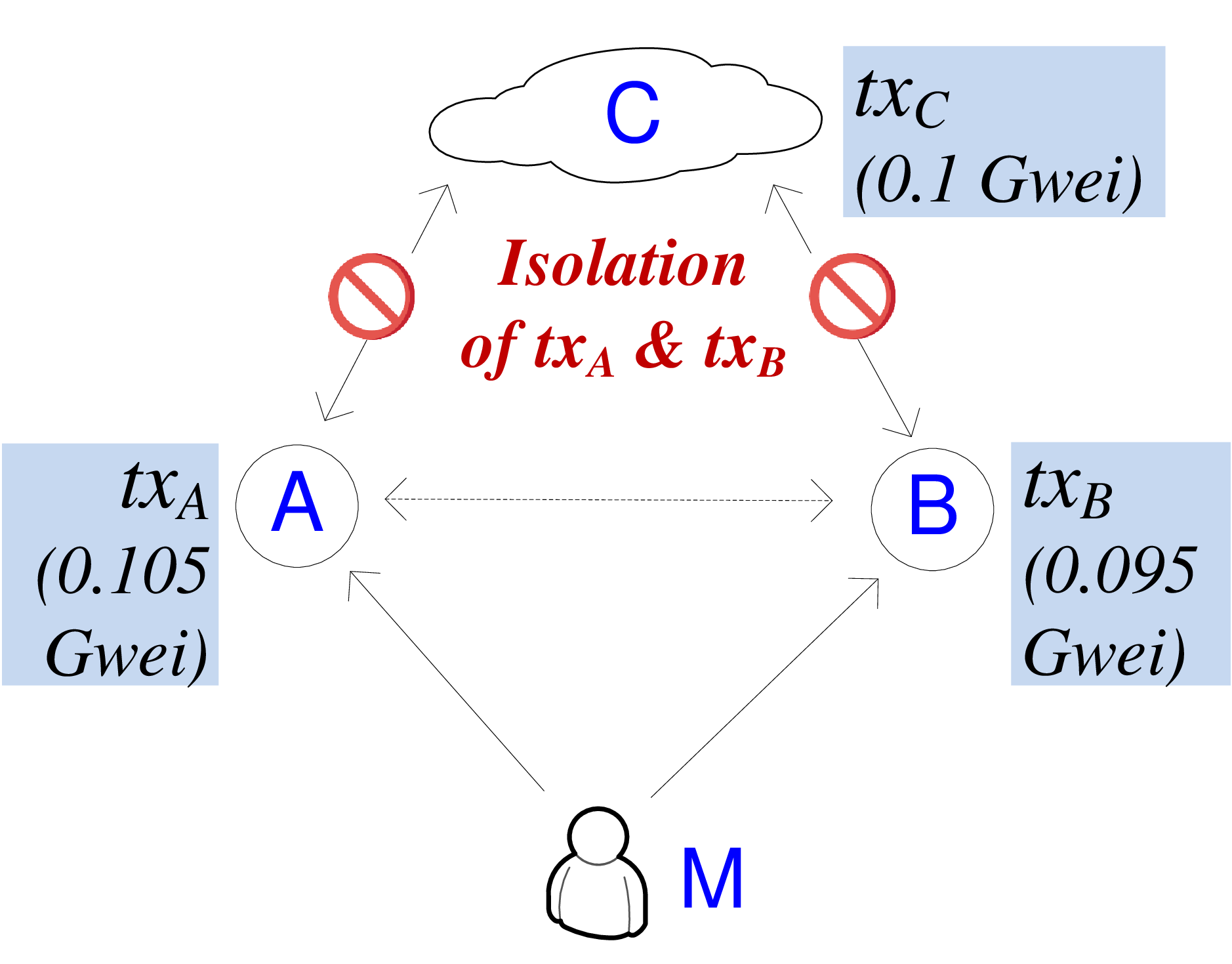}%
  \label{fig:example}
    }%
  \end{center}
  \caption{\sysname's measurement primitive: Running $measureOneLink$ with $Y=0.1$ Gwei, $Z=5120$, $R=10\%$, $U=1$}
\end{figure*}

\subsection{Measurement Primitive}
\label{sec:measure:serial}
We consider the basic system model consisting of a measurement node $M$, target node $A$, target node $B$ and the rest of Ethereum network denoted by node(s) $C$. The measurement primitive's goal is to detect one link, that is, whether Node $A$ and $B$ are actively connected in the Ethereum network. Note that this model assumes a strongly connected Ethereum network without network partition.

\noindent{\bf Mechanism}:
We denote our measurement primitive by $measureOneLink(A,B,X,Y,Z,R,U)$, which are parameterized by target nodes $A$ and $B$, target nodes' \texttt{mempool} behavior $R$/$U$ (recall Table~\ref{tab:notate}) and 
$X$/$Y$/$Z$ that will be described below. 
As depicted in Figure~\ref{fig:workflow}, the measurement primitive works in four steps:

\begin{itemize}
\item[\ballnumber{\scriptsize \footnotesize 1}]
Node $M$ sends a pending transaction $tx_C$ with Gas price $Y$ Gwei\footnote{One Gwei equals $10^{-9}$ Ether.} to A, and waits for $X$ seconds (e.g., $X=10$ in our study as will be described) for $tx_C$ to be propagated to other nodes including node $B$. Setting $Y$ at a low Gas price is intended to slow down or even prevent the inclusion of $tx_C$ in the next block (recall Ethereum nodes decide which transactions to be included in the next block based on Gas/Gas price).
\item[\ballnumber{\scriptsize \footnotesize 2}]
Node $M$ sends to Node $B$ $Z$ future transactions $\{tx_{O1},tx_{O2},...tx_{OZ}\}$ at Gas price $(1+R)\cdot{}Y$ Gwei. These future transactions are uniformly sent from $\frac{Z}{U}$ accounts (i.e., there are $U$ future transactions sent from each account).
Immediately after that, Node $M$ sends a transaction $tx_B$ at Gas price $(1-0.5R)\cdot{}Y$ Gwei to Node $B$.
Transaction $tx_B$ has the same nonce with $tx_C$.
\item[\ballnumber{\scriptsize \footnotesize 3}]
Node $M$ sends to Node $A$ $Z$ future transactions $\{tx_{O1},tx_{O2},...tx_{OZ}\}$ which are at Gas price $(1+R)\cdot{}Y$ Gwei and sent from $\frac{Z}{U}$ accounts. Immediately after that, Node $M$ sends a transaction $tx_A$ at Gas price $(1+0.5R)\cdot{}Y$ Gwei to Node $A$.
Transaction $tx_A$ has the same nonce with $tx_C$.

The purpose of the future transactions is to fill up the \texttt{mempool} on Nodes $A$ (and $B$), to evict $tx_C$ there, and to make room for $tx_A$ ($tx_B$) of the same nonce to $tx_C$.
\item[\ballnumber{\scriptsize \footnotesize 4}]
Node $M$ checks if it receives $tx_A$ from Node $B$. If so, it draws the conclusion that Node $A$ is a neighbor of Node $B$, as will be analyzed in \S~\ref{sec:analysis}.
\end{itemize}

\noindent{\bf Configuration of $R$/$U$}:
Parameters of the $measureOneLink$ primitive are configured as follows: On a target Ethereum client, parameters $R$/$U$ will be set at the client's value as in Table~\ref{tab:measure:diffclients}. Here, note that Nethermind and Aleth are not measure-able by \sysname due to their zero-value $R$ which is also flawed as explained before. Besu has an infinite large value of $U$, and Geth has a fairly large $U$. In these two cases, only one account is used to send the future transactions $\{tx_{O}\}$. Geth/Parity have non-zero $P$, which are fairly small compared with their \texttt{mempool} capability $L$. The working of measureOneLink requires the following condition:
{\it The number of pending transactions in the measured \texttt{mempool} should remain larger than $P$ in the entire process of measurement.} We verify that this condition holds on the mainnet for all Ethereum clients' $P$ and $L$.

\noindent{\bf Configuration of $X$}:
Parameter $X$, which is the time period that Step \ballnumber{\scriptsize \footnotesize 1} waits, is set to be large enough so that transaction $tx_C$ is propagated to all nodes in the network. In order to obtain a proper value of $X$ in an Ethereum network, we conduct a test by running several local nodes (e.g., $11$ nodes in our study) and joining them to the Ethereum network. Among the $11$ nodes, there are no direct connections. During the test, we send a transaction through one node, wait for $X'$ seconds and observe the presence of the transaction on the other $10$ nodes. We conduct a series of such tests with varying $X'$es to obtain such a $X'=X$ that with $99.9\%$ chances, the transaction is present on the $10$ nodes after $X$ seconds. 

The four steps occur in order. That is, Step \ballnumber{\scriptsize 1} occurs $X$ seconds before Step \ballnumber{\scriptsize 2}, which finishes before Step \ballnumber{\scriptsize 3} starts, which is before Step \ballnumber{\scriptsize 4}.
Timing and ordering are essential to the success of our measurement method, as is analyzed below. 

\subsubsection{Correctness Analysis}
\label{sec:analysis}

We analyze the correctness of the measurement primitive ($measureOneLink$):

$10$ seconds after Step \ballnumber{\scriptsize 1}, Transaction $tx_C$ is propagated to the entire Ethereum network and it is stored in all nodes' \texttt{mempool}s including Nodes $A$ and $B$. 

During \ballnumber{\scriptsize 2}, when Node $B$ receives $Z$ future transactions $tx_O$s, its \texttt{mempool} becomes full. Based on the eviction policy in Table~\ref{tab:measure:diffclients}, adding a new transaction to a full \texttt{mempool} triggers evicting the transaction with the lowest Gas price. 
Assuming Gas price $Y$ Gwei is low enough (we will describe how to set $Y$ next), transaction $tx_C$ at $Y$ Gwei will be evicted on Node $B$. 
Then, without $tx_C$, transaction $tx_B$ is stored in Node $B$'s \texttt{mempool}. In other words, Step \ballnumber{\scriptsize 2} replaces $tx_C$ with $tx_B$ on Node $B$.
Note that in the process, future transactions $\{tx_O\}$ are not propagated, thus $C$ still stores $tx_C$.

Note that after the arrival of $\{tx_{O}\}$ but before $tx_B$, there are chances that certain Nodes $C$ can propagate $tx_C$ back to Node $B$, which, if occurs, would invalidate the efforts of $\{tx_{O}\}$ and leave $tx_B$ unable to replace (the re-propagated) $tx_C$ on $B$. In \sysname, the actual chance of this event is very low and the reason is two-fold: 1) \ballnumber{\scriptsize 2} waits long enough ($10$ seconds) after \ballnumber{\scriptsize 1} to start and 2) $tx_B$ is propagated immediately after \{$tx_{O}\}$. {In addition, in our local validation experiment (in \S~\ref{sec:measure:validation}), we don't observe the occurrence of the event.}

By a similar analysis, Step \ballnumber{\scriptsize 3} can replace $tx_C$ with $tx_A$ on Node $A$. 

Now, we have established that after Steps \ballnumber{\scriptsize 1},\ballnumber{\scriptsize 2} and \ballnumber{\scriptsize 3}, Node $A$ stores $tx_A$, Node $B$ stores $tx_B$ and Nodes $C$ store $tx_C$. The snapshot of our measurement system at this timing is illustrated in Figure~\ref{fig:example}.

We consider two cases: Case 1) $A$ and $B$ are directly connected. In this case, $A$ will propagate $tx_A$ to $B$, which will replace $tx_B$ because of $tx_A$'s $R$  (e.g., $10\%$ for Geth) higher Gas price than $tx_B$. In this case, $A$ will also propagate $tx_A$ to $C$, which however will not replace $tx_C$ as $tx_A$'s Gas price is lower than $R$ (e.g., $10\%$) of $tx_C$'s price. 
The property that $tx_A$ is stored only on Node $A$ and cannot be propagated through Nodes $C$ is called isolation. That is, $tx_A$ is isolated on Node $A$. Thus, after a sufficient delay for propagation from $A$ to B, $M$ can receive $tx_A$ from Node $B$.

Case 2) $A$ and $B$ are not connected. In this case, $A$ propagates $tx_A$ only to Node $C$. As analyzed, $tx_A$ cannot replace $tx_C$ on Node $C$ because of insufficient Gas price. Also, Node $C$'s $tx_C$ cannot replace $tx_B$ on Node $B$. Thus,  $tx_B$ stays on Node $B$, and $M$ does not receive $tx_A$ from Node $B$.  

{To ensure correctness, \sysname requires that the \texttt{mempool} on the two measured nodes, namely nodes $A$ and $B$, are full. This condition holds quite commonly in Ethereum mainnet, as observed in our measurement study ($99\%$ of the time during our mainnet measurement, the measurement node's \texttt{mempool} is full). 
}

\label{sec:gasprice}
\noindent{\bf Configuration of $Y$/$Z$}:
Pending transactions like $tx_C$ should stay in the \texttt{mempool} of Nodes $C$, in such a way that they are not included in the next block or be evicted.
To do so, the Gas price of $tx_C$ should be low enough so that it will not be included in the next block, and at the same time, be high enough to avoid eviction by incoming transactions. 
To estimate a proper Gas price in the presence of current transactions, we rank all pending transactions in the \texttt{mempool} of Node $M$ by their Gas prices, and use the median Gas price for $tx_C$. In actual measurement studies, the value of $Y$ varies from testnets and at different times. We apply the estimation method before every measurement study and obtain $Y$ dynamically.

{\subsubsection{Cost Analysis}

The cost of running $measureOneLink$ comes from the pending transactions sent (i.e., either $tx_A$ or $tx_C$), assuming their inclusion in the blockchain. In practice, whether these two transactions are included is not deterministic and depends on the state of the miners' \texttt{mempool}. Also, note that the future transactions $tx_O$ sent during the measurement are guaranteed not to be included in the testnets and mainnet, thus incurring no costs.
}

\subsubsection{Improving Result Recall}

Based on the above analysis, the \sysname guarantees that any tested connection is a true positive (i.e., no false positives) but may miss the detection of a connection (i.e., false negative may exist). In other words, the $100\%$ result precision is guaranteed by the protocol but not for the recall. 
{Note that $100\%$ precision/recall means no false positive/no true negative in the measurement result.}
In the following, we present several heuristics to improve the result recalls in practice.

A passive method to improve the result recall is to repeat the measurements multiple times and use the union of the results. This passive method has limited applicability if the false negative is caused by the non-default setting on the remote Geth node being measured. In the following, we propose a proactive method to improve the recall.

{{\bf Handling node-specific configurations by pre-processing}:
In Ethereum networks, client configurations (e.g., on \texttt{mempool}) are specific to nodes. This is evident in our field experience where the \texttt{mempool} capacities (i.e., $L$) differ across nodes. Using the same value of $L$ when measuring different nodes can lead to incorrect results. 

To solve the problem,} we add a pre-processing phase: Before the measurement, we can launch a speculative $B'$ node locally and use it to connect all other nodes in the network. For each other node, say $A'$, we then run \sysname between $A'$ and $B'$. Because the local node $B'$ is under our control and its actual neighbors can be known (by sending \texttt{peer\_list} RPC queries), we compare the measurement result with the ground truth. If there is a false negative, it implies the remote node $A'$ has some non-default setting on its node (e.g., use a \texttt{mempool} larger than the default $Z$). We then increase the \texttt{mempool} size in additional pre-processing measurements to discover a proper setting of the \texttt{mempool}. The result of the pre-processing can help guide the actual measurement to use a ``right'' parameter on the connections  involving node $A'$.

\subsection{Parallel Measurement Framework}
\label{sec:measure:par}

We previously described the primitive of measuring one connection between a source and a sink node. To measure a network, a native schedule is to serially run the pair-wise primitive over all possible pairs, which however incurs a long measurement time in the case of large networks and is not a scalable method. For time efficiency, we propose a parallel schedule that decomposes the set of all possible pairwise connections into subsets and measures all connections within each subset in parallel. In the following, we first describe the parallel measurement primitive (\S~\ref{sec:measure:par:primitive}) and then the schedule that measures the entire network in parallel by repeatedly using the primitives (\S~\ref{sec:measure:par:schedule}). 

\subsubsection{Parallel Measurement Primitive}
\label{sec:measure:par:primitive}

{
We consider a pair of nodes whose connectivity is measured consist of a source node and a sink node. For instance, in Figure~\ref{fig:example}, node $A$ is a source node and node $B$ is a sink node. In a parallel measurement, we consider measuring the connectivity between not one pair of source and sink nodes, but multiple such pairs. Specifically, suppose
} there are $p$ ``source'' nodes $A_1,A_2,...A_k,...A_p$ and $q$ ``sink'' nodes $B_1,B_2,...,B_l,...,B_q$; note $k$ ($l$) is the index of a source (sink) node. In this bipartite graph, there are a total of $p\cdot{}q$ possible edges from a source to a sink. The objective here is to measure $r$ specified edges out of the $p\cdot{}q$ ones. 
We denote by $sink(k,j)$ a sink node which is the $j$-th neighbor of a source node $A_k$.
Then, the edge between $A_k$ and $sink(k,j)$ is ``indexed'' by $(k,j)$.
Initially, assume there are sufficient funds set up in $r$ Externally Owned Accounts (or EOAs).

\begin{itemize}
\item[\ballnumber{\scriptsize p1}]
Node $M$ sends a total of $r$ transactions $\{tx_{C(k,l)}\}$ and propagates them to the Ethereum network. Any two different transactions are sent from different EOAs.

\item[\ballnumber{\scriptsize p2}]
To each Node $A_k$, Node $M$ 1) first sends $Z$ (e.g., 5120 for Geth) future transactions $tx_{F}$'s followed immediately by 2) sending $\{...,tx_{C(k-1,q_{k-1})},tx_{C(k+1,1)},...\}$. 3) It then sends $\{tx_{A(k,1)},...tx_{A(k,q_k)}\}$. Here, $tx_{A(k,i)}$ spends the same account with $tx_{C(k,i)}$ and its Gas is priced at $1.05Y$ Gwei. After \ballnumber{\scriptsize p2}, $tx_{C(k,i)}$ on Node $Ai$ is replaced by $tx_{A(k,i)}$, while other $tx_C$'s stay.

It is noteworthy that after \ballnumber{\scriptsize p2}, Node $M$ checks whether $tx_{A(k,\cdot)}$ are actually stored on Node $A_k$. It proceeds only if the checked condition holds. Node $M$ carries out the check by observing if $tx_{C(k,\cdot)}$ is propagated from Node $A_k$ before waiting for a timeout.

\begin{figure*}[!ht]
  \begin{center}
  \subfloat[Network snapshot before Step p4]{%
  \includegraphics[width=0.35\textwidth]{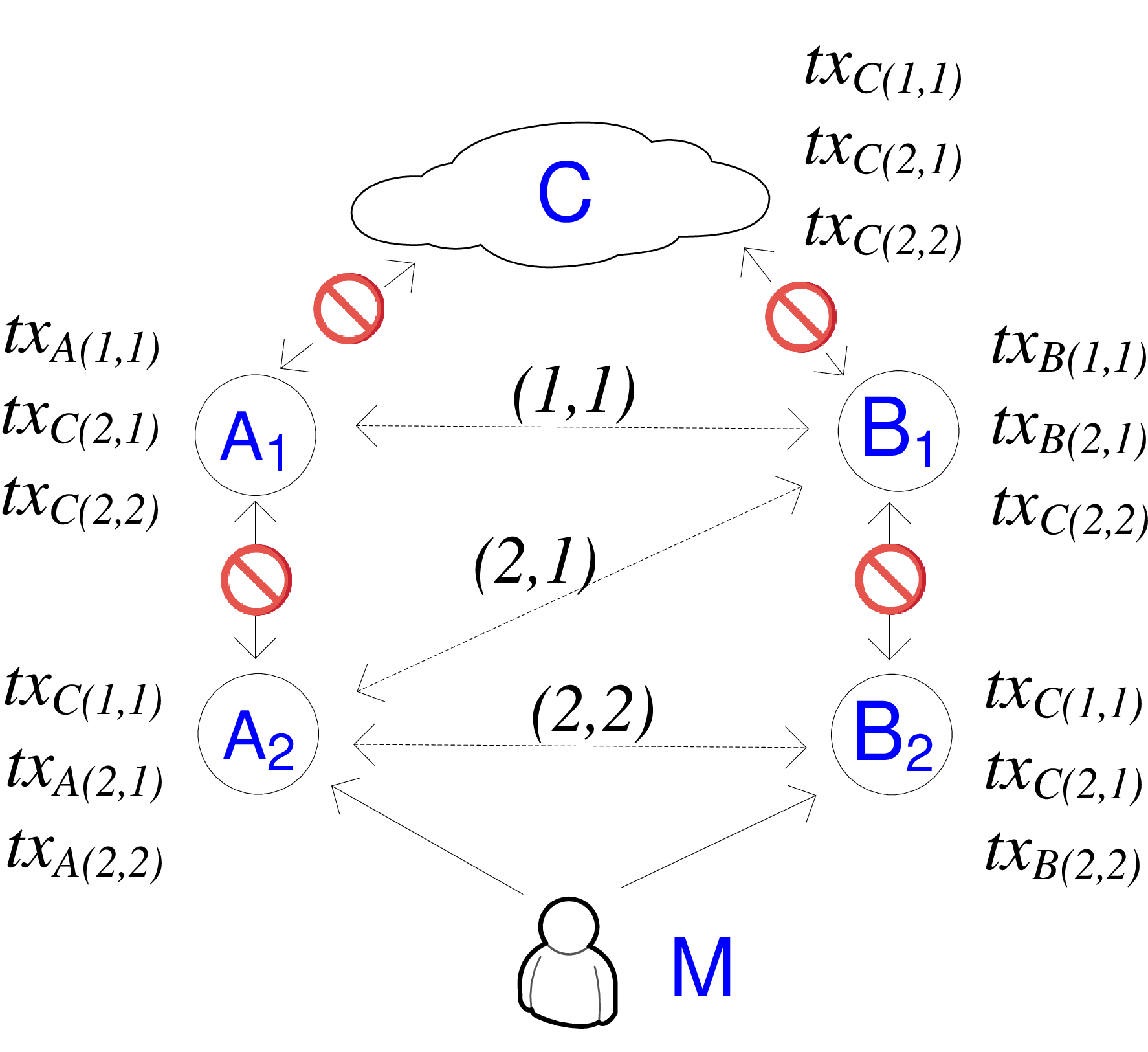}%
  \label{fig:parallel}
  }
  \hspace{1.5in}
  \subfloat[Parallel-measurement schedule with an example network of $8$ nodes]{%
  \includegraphics[width=0.275\textwidth]{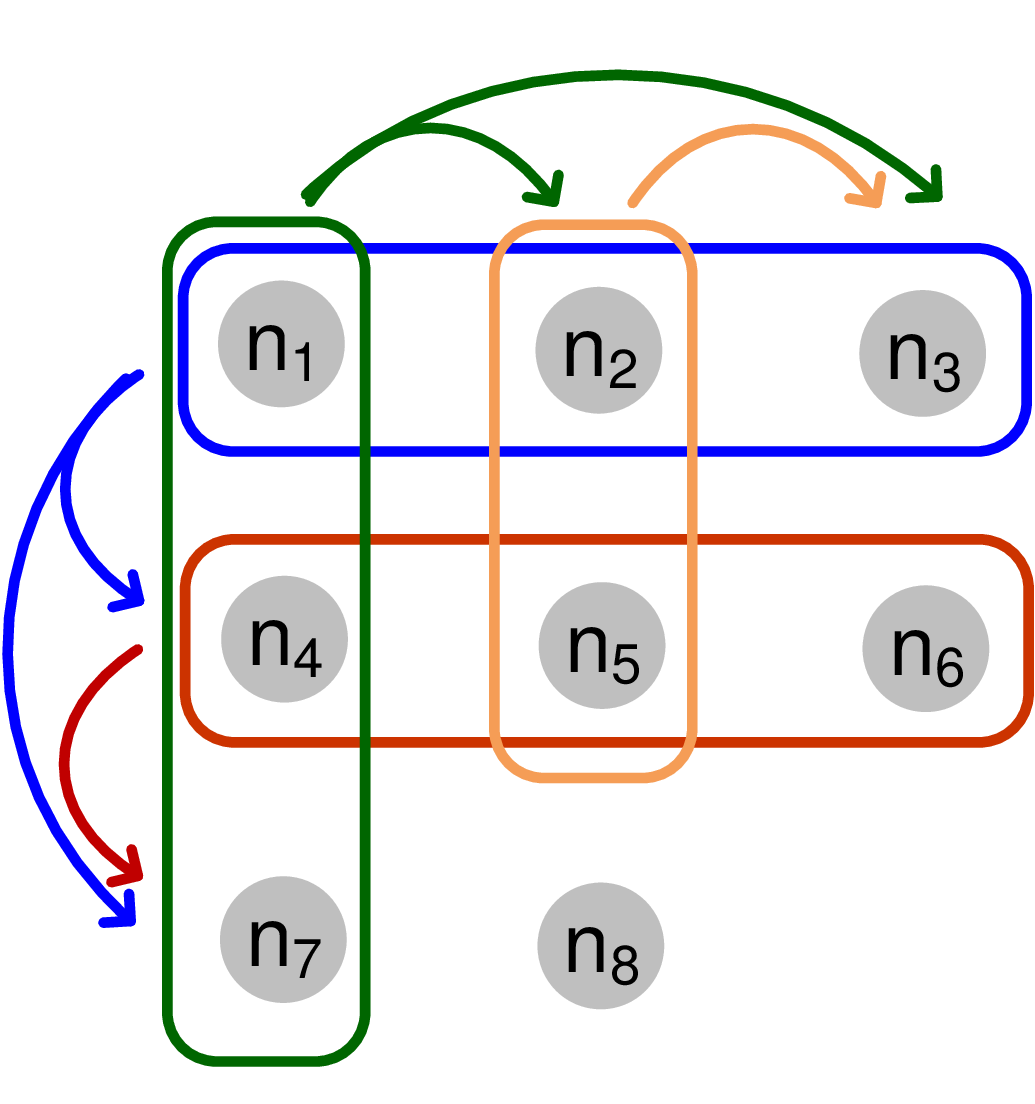}%
  \label{fig:parallel:example}
  }
  \end{center}
  \caption{\sysname's parallel measurement protocol; in Figure~\ref{fig:parallel:example}, the four colors represent four $measurePar$ iterations where a rectangle with rounded angles is the selected nodes $A$'s in the measurement and the arcs are the collection of edges being measured. For instance, the blue iteration is $measurePar(\{n_1,n_2,n_3\},\{n_4,n_5,n_6,n_7,n_8\},\{\emptyset\})$.}
\end{figure*}

\item[\ballnumber{\scriptsize p3}]
Node $M$ sends to each Node $B_l$ $Z$ future transactions $tx_{F}$'s followed immediately by $r$ transactions where the $i$-th transaction is a $tx_B$ transaction (whose Gas is $95\%$ of that of a $tx_C$ transaction)  if the $i$-th edge's sink is Node $B_l$, and otherwise, is a $tx_c$. 

\item[\ballnumber{\scriptsize p4}]
For edge connecting $A_k$ and $B_l$, Node $M$ checks if $tx_{A(k,j)}$ (note that $sink(k,j)=l$) is present on Node $B_l$. If so, ${A_k}$ and $B_l$ are neighbors.
\end{itemize}

{Note that Ethereum clients, including both Geth and Parity, limit the number of future transactions in their \texttt{mempool}. In our parallel measurements, we ensure the group size is much smaller than the limit of future transactions, which further ensures the measurement correctness, since all measurement transactions will be admitted and stored on the participant nodes.
}

{\bf Example}: We use an example to illustrate the parallel measurement protocol. Among two sources $A_1,A_2$ and two sinks $B_1,B_2$, assume it measures the connections on three edges, that is, $\langle{}A_1,B_1\rangle{}, \langle{}A_2,B_1\rangle{}, \langle{}A_2,B_2\rangle{}$. Figure~\ref{fig:parallel} depicts the snapshot of exercising our parallel measure method right after \ballnumber{\scriptsize p3}.

{\bf Ensuring isolation}: As in the case of measurement primitive, isolation is critical to the success of our measurement method. In the parallel setting, a Node $A$ needs to prevent propagating the $tx_A$ transactions to Nodes $B$'s via not only Nodes $C$'s but also other Nodes $A$'s and Nodes $B$'s. For instance, in the example above, when measuring the connection between Node $A_2$ and $B_1$, it needs to ensure that $tx_A(2,1)$ is not propagated to Node $B_1$ via Node $A_1$ or $B_2$. This is guaranteed by our measurement method because Nodes $B_2$ and $A_1$ store $tx_C$ transactions and can be treated as a $C$ node when measuring the connection between Node $A_2$ and $B_1$.

\subsubsection{Parallel Measurement Schedule}
\label{sec:measure:par:schedule}

Given a network of nodes $\{n_1,n_2,...,n_N\}$, we partition the nodes into $N/K$ groups where each group is of $K$ distinct nodes; for instance, the $i$-th group ($i$ starting from $0$) is of nodes $\{n_{i*K+1},n_{i*K+2},...n_{i*K+K-1}\}$.

We schedule the network measurement in the two rounds: The first round runs $N/K$ iterations, where each iteration measures the edges between group $i$ and the rest of the network. The second round measures the edges within a group.

To be more specific, we denote the parallel measurement primitive described in \S~\ref{sec:measure:par:primitive} by $measurePar(\{A_i\},\{B_i\},\{C\})$. 1) Given the $i$-th group, the first round calls $measurePar(\{n_{i*K+1},n_{i*K+2},...n_{i*K+K-1}\},\{n_{1},...,n_{i*K-1},n_{i*K+K},$ $...n_{N}\},\{\emptyset\})$, where A is the $i$-th group, B is the rest of the blockchain network, and C is empty. Each of these iterations sets a goal to measure $K\cdot{}(N-K)$ possible edges.

2) The second round measures the edges within groups. Specifically, given a group, it maps the first half of nodes as $A$ and the other half as Nodes $B$. An iteration measures intra-group edges for all groups. It then applies the same splitting respectively for the first and second half of the group. In other words, the next iteration measures the intra-group edges for half of the original groups. This process repeats until the group size reaches $2$. 

{\bf Example}: Suppose $N=8$ and $K=3$. The parallel schedule is of two rounds, each of two iterations, as illustrated by the four curved rectangles (with different colors) in Figure~\ref{fig:parallel:example}.

The first round runs the following two iterations: 
$measurePar(\{n_1,n_2,n_3\},\{n_4,n_5,n_6,n_7,n_8\},\{\emptyset\})$ which measures the $3*5=15$ edges across node group $\{n_1,n_2,n_3\}$ and group $\{n_4,n_5,n_6,n_7,n_8\}$. This is visualized by the horizontal rectangle in blue in the figure.
The second iteration is $measurePar(\{n_4,n_5,n_6\},\{n_7,n_8\},\{n_1,n_2,n_3\})$ which measures all $3*2=6$ edges and is visualized by the horizontal rectangle in red in the figure.

The second round runs another two iterations: $measurePar(\{n_1,n_4,n_7\},\{n_2,n_3,n_5,n_6,n_8\},\{\emptyset\})$ which measures $5$ edges across groups (i.e., edges $(n_1,n_2),(n_1,n_3),(n_4,n_5),(n_4,n_6),(n_7,n_8)$) by the vertical rectangle in green, and $measurePar(\{n_2,n_5\},\{n_3,n_6\},\{n_1,n_4,n_7,n_8\})$ which measures 2 edges (i.e., $(n_2,n_3),(n_5,n_6)$) by the vertical rectangle in orange. 

\ignore{
$measurePar(\{n_1,n_2,n_3\},\{n_4,n_5,n_6,n_7,n_8\},\{\emptyset\})$ which measures all $3*5=15$ edges, $measurePar(\{n_4,n_5,n_6\},\{n_1,n_2,n_3,n_7,n_8\},\{\emptyset\})$ which measures all $15$ edges, and $measurePar(\{n_7,n_8\},\{n_1,n_2,n_3,n_4,n_5,n_6\},\{\emptyset\})$ which measures $2*6=12$ edges.
}

{\bf Complexity Analysis and Configuration of $K$}: On the measurement of a network of $N$ nodes with a group of size $K$, the total number of iterations is $\frac{N}{K}+\log{K}$ where the first round runs $\frac{N}{K}$ iterations and the second round runs $\log{K}$ iterations. Roughly, the number of iterations decreases with increasing $K$. However, making the value of $K$ too large would lead to the overflow of \texttt{mempool} as it generates $K*(N-K)$ transactions in each iteration. In practice, an Ethereum node's \texttt{mempool} has a capacity of $5120$ transactions and to bound the interference, we only use no more than $2000$ transaction slots in the \texttt{mempool}. For an Ethereum network of 500 nodes, such as Ropsten, we use $K=2000/500=4$ which results in a total of $500/4+\log{4}$=$127$ iterations.

\section{Measurement Results}
\label{sec:results}

Initially, we run a measurement node $M$ that joins an Ethereum network, such as the Ropsten testnet.
The measurement node $M$ is set up without bounds on its neighbors, so it can be connected to the majority of the network.

\subsection{Measurement Validation}
\label{sec:measure:validation}
The correct functioning of \sysname relies on several factors that may vary in a deployed Ethereum network. For instance, \sysname assumes the default \texttt{mempool} size on Geth nodes (i.e., 5120) that may not hold if an Ethereum node is configured with a different \texttt{mempool} size. The variance would introduce false negatives into \sysname results and affect the recall. In this subsection, we validate the \sysname results by evaluating/estimating the result recall. Here, we use the ``external'' experiment setup; in Appendix~\ref{sec:validate:local}, we use a fully local setup to conduct additional validation study.

{\bf Experiment setup}: In addition to the measurement node $M$, we set up a local machine to play node $B$; the node joins the testnet being measured (e.g., Ropsten) and is configured with a number (e.g., $5000$) larger than the size of testnet. After staying online for 12 hours in Ropsten, node $B$ connects to 520 nodes, among which 471 nodes run Geth clients. The setup here is external as nodes $A$ and $B$ join a remote Ethereum network.

{\bf Validating measurement primitive ($measureOneLink$)}: We then iterate through the 471 nodes, selecting each node as node $A$ to measure the connection between $B$ and $A$ using an unmodified \sysname. In each iteration, the connection is measured three times. 
{When running the measurement primitive, we verify that $tx_C$ is evicted from nodes $A$ and $B$. This is done by turning on the RPC interface and sending an \texttt{eth\_getTransactionByHash} query to it.}
The final result is positive (i.e., there is a connection) if any of the three measurements is positive. For each unit experiment, we report the number of positive connections \sysname can detect and from there calculate the recall.


\begin{figure}
\centering
    \subfloat[Recall with \sysname sending increasing number of future transactions.]{%
  \includegraphics[width=0.225\textwidth]{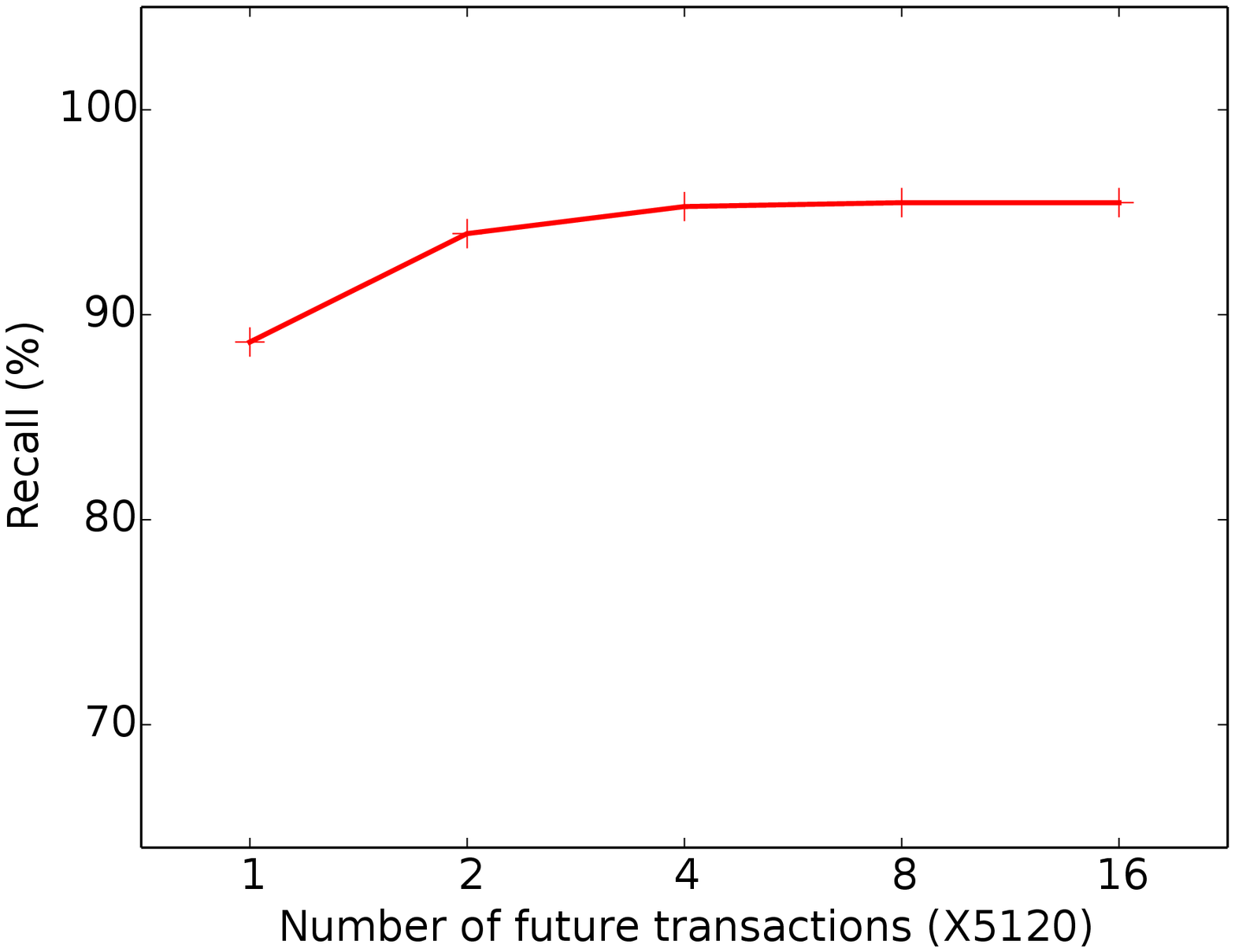}%
  \label{fig:recall:future}} \hspace{0.1in}
    \subfloat[Precision and recall with increasing group size in parallel measurement.]{%
  \includegraphics[width=0.225\textwidth]{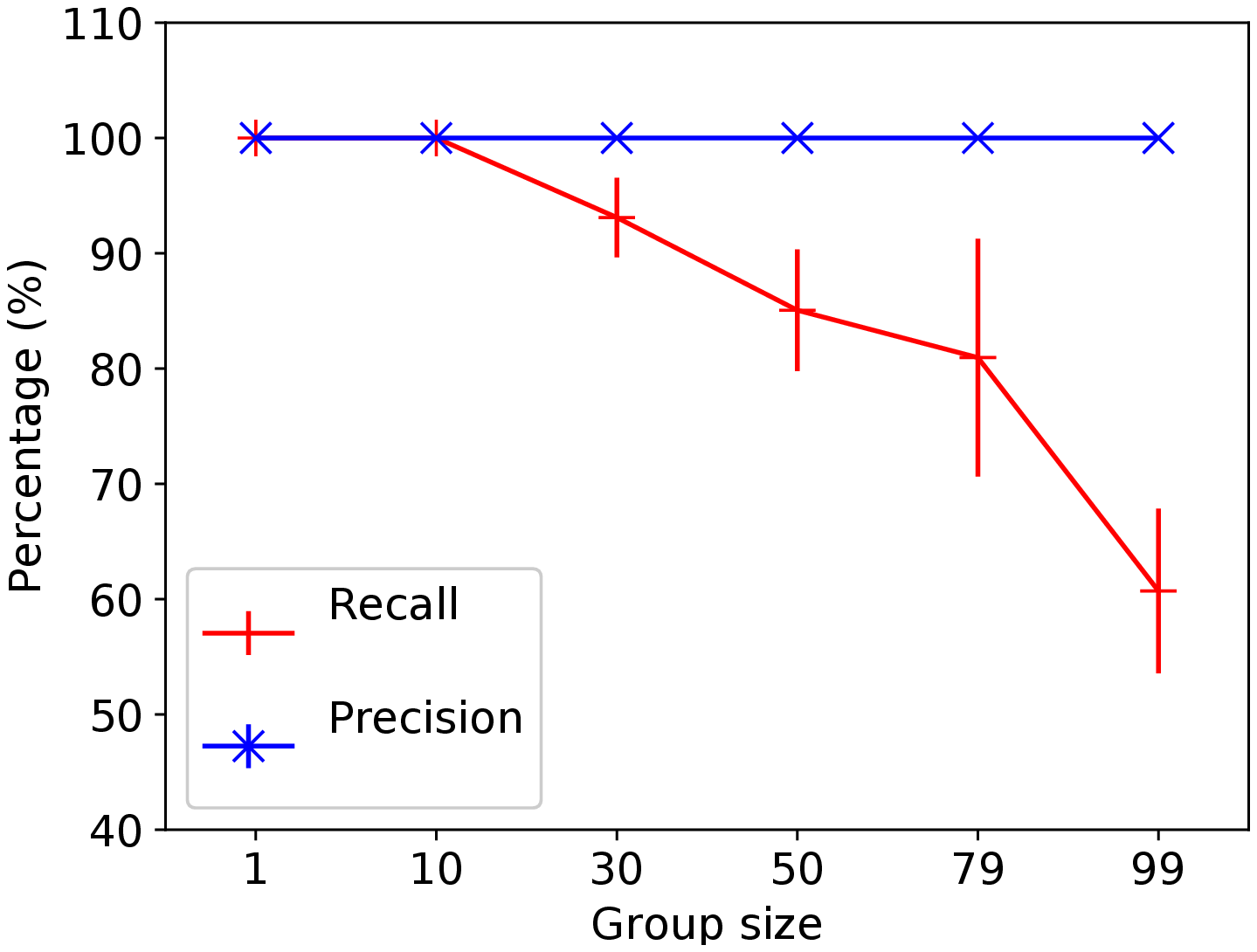}%
  \label{fig:parallel:eval}}
  \caption{Measurement validation results}
\end{figure}

We increase the number of future transactions sent in \sysname and measure the recall using the validation method above. The results are shown in Figure~\ref{fig:recall:future}. With the increasing number of future transactions, the recall of \sysname grows from $84\%$ to $97\%$. An implication here is that even with a large number of future transactions, \sysname does not reach 100\% recall. We suspect the following culprits: 1) The remote node is configured with a custom \texttt{mempool} size much larger than the default 5120. 2) The node is configured with a custom Gas price threshold other than the default 10\%; this threshold determines the \texttt{mempool}'s transaction replacement policy. 3) There are nodes who join Ropsten testnet but do not participate in forwarding transactions, preventing $tx_A$ being propagated.

{\bf Validating parallel method ($measurePar$}: In the same experiment we then validate \sysname's parallel measurement method.

Recall parallel \sysname is parameterized with $p$ and $q$. In this experiment, we use $q=1$ and vary $p$ (referred to as the group size), that is, a node $B'$ and $p$ nodes $A$'s in a parallel measurement. $p$ is varied between $1$ to $99$. 

{\color{black}
Specifically, we set up a new node $B'$ with the default $50$ active neighbors and join the Ropsten testnet. It turned out its $35$ active neighbors run Geth. We then serially measure the $35$ neighbors, which successfully detects $29$ neighbors. When running validation of the parallel method, we need to choose $p$ nodes $A$'s. When $p\leq{29}$, we choose a subset of the $29$ active neighbors of node $B'$ to play nodes $A$'s. When $p>29$, we choose the $29$ neighbors of node $B'$, as well as the nodes that do not have connections with node $B'$, to be nodes $A$'s.
}

For each group size, we run the parallel measurement three times and report a positive result if any of the three returns a positive result. The results are presented in Figure~\ref{fig:parallel:eval}. 
The precision is always 100\%. The recall is initially 100\% until the group size is larger than $29$. It then decreases as the group grows larger. For a group of 99 nodes, the recall is about 60\%. The reason for a non-100\% recall under a large group is that \sysname does not guarantee isolation among nodes $\{A\}$, and a larger group increases the chance of non-isolation/interference among nodes $\{A\}$.

\begin{figure}[!ht]
  \begin{center}
\includegraphics[width=0.35\textwidth]{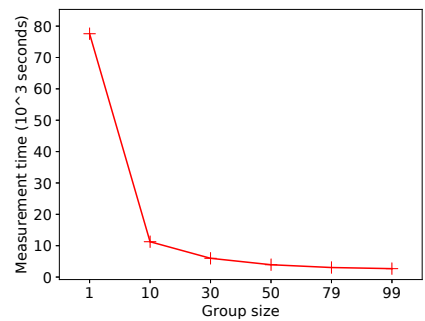}
  \end{center}
  \caption{Speedup of \sysname's parallel measurement over the serial measurement}
  \label{fig:timespeedup}
\end{figure}

{\bf Measurement speedup of the parallel method}: We also report the time of measuring the same group of nodes with varying group size, with the purpose to evaluate possible speedup by the parallel measurement over the serial one. In a similar experiment setup, the measurement target is a group of $100$ nodes.  With about $4950$ edges detected, the measurement times are reported in Figure~\ref{fig:timespeedup}. It can be seen that as the group size $K$ increases, the time to measure the same group of nodes (as in the previous experiment) decreases significantly. For instance, with a group size $K=30$, the measurement time is reduced by an order of magnitude (about $10\times$ times).

\subsection{Testnet Measurement Results}
\subsubsection{Ropsten Results}

\begin{figure}[!ht]
  \begin{center}
    \includegraphics[width=0.3\textwidth]{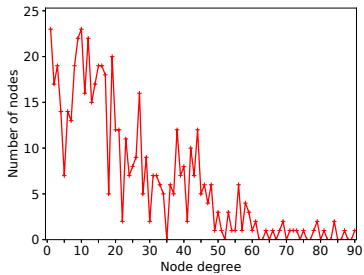}%
  \end{center}
  \caption{Node degree distribution in Ropsten}
  \label{fig:ropsten:degree}
\end{figure}

\hspace{0.05in}
\begin{table}[!htbp] 
\caption{Graph properties of the Ropsten testnet}
\label{tab:graphproperties:ropsten}\centering{\footnotesize
\begin{tabularx}{0.45\textwidth}{ |X|X|X|l|X| }
  \hline
&
Measured Ropsten & ER ($n=588$, $m=7496$) & CM & BA ($n=588$, $l'=26$)
\\ \hline
Diameter & 5 & 3.0 & 5.2 & 3.0
\\ \hline
Periphery size & 36 & 293.5 & 24.9 & 509.4 
\\ \hline
Radius & 3 & 3.0 & 3.0 & 2.0
\\ \hline
Center size & 36 & 293.5 & 51.7 & 78.6   
\\ \hline
Eccentricity & 4.037 & 3.0 & 3.98 & 2.87
\\ \hline
Clustering coefficient & 0.207 & 0.044 & 0.139 & 0.159   
\\ \hline
Transitivity & 0.127 & 0.044 & 0.122 & 0.156 
\\ \hline
Degree assortativity & -0.1517 & 0.0026 & -0.0664 & -0.0181
\\ \hline
Clique number & 60.75 & 250.3 & 557.4 & 50.6
\\ \hline
Modularity & 0.0605 & 0.161 & 0.152 & 0.102 \\ \hline
\end{tabularx}
}
\end{table}

\begin{table}[!htbp] 
\caption{Detected communities in Ropsten testnet}
\label{tab:ropsten:comm}\centering{\small
\begin{tabularx}{0.45\textwidth}{ |X|l|X|X| }
 \hline
Community index & No. of nodes & Intra-comm. edges (density) & Inter-comm. edges
 \\ \hline
1 & 92 & 423 (10\%) & 1547
 \\ \hline
2 & 142 & 603 (6\%) & 1612
 \\ \hline
3 & 107 & 548 (9.7\%) & 1827
 \\ \hline
4 & 84 & 391 (11\%) & 1505
 \\ \hline
5 & 75 & 379 (14\%) & 1704
 \\ \hline
6 & 51 & 127 (10\%) & 773
 \\ \hline
7 & 37 & 121 (18\%) & 840
 \\ \hline
\end{tabularx}
}
\end{table}

We first conducted a measurement study on Ropsten testnet. We use parallel measurement method with parameter $K=60$. In particular, the testnet is under loaded and there are not sufficient ``background'' transactions in \texttt{mempool}s. We tried to apply \sysname, as is, to measure Ropsten and found that however low Gas price we set for $tx_C$ (recall Step \ballnumber{1} in \sysname), the transaction will always be included in the next block, leaving no time for accurate measurement. To overcome this problem, we launch another node that sends a number of ``background'' transactions (from a different account than $tx_C$). This effectively populate an operating \texttt{mempool} and helped $tx_C$ stay in a \texttt{mempool} for long enough during the measurement period. We encounter the same situation when measuring Goerli and use the same trick here.
{Note that more than $95\%$ of peer nodes our supernodes initially connect to stay connected throughout the measurement period.

In the testnet, a target node may run a non-default setting in which they forward future transactions, invalidating the assumption made in \sysname. Such a custom node is avoided in our measurement study as follows: In the pre-processing, one can launch an additional monitor node (to the measurement node) to connect to the target node one. The measurement node then sends a future transaction to the target node. If the monitor node observes the future transaction from the target node, the target node is removed from the measurement. Besides, the pre-processing phase in \sysname also avoids unresponsive nodes.
}

We present a snapshot of the Ropsten testnet taken on Oct. 13, 2020. The precision of the measurement result is $100\%$ and recall is $88\%$ (under group size $K=60$), using a validation method described above. The network contains 588 (Geth) nodes and 7496 edges among them. 
This result has the test node and its edges excluded.
The degree distribution is plotted in Figure~\ref{fig:ropsten:degree}. Most nodes have a degree between $1$ and $60$: Particularly, $4\%$ of nodes have degree 10, another 4\% have degree 1 and another $4\%$ have degree 12. Omitted in the figure are ten nodes with degree between 90 and 200. This result shows that degrees by active links are much smaller than the default number of inactive neighbors ($250$). 

Table~\ref{tab:graphproperties:ropsten} summarizes the characteristics of the measured testnet in terms of distances, assortativity, clustering and community structure. 1) For distances, the network diameter, defined as the maximal distance between any pair of nodes, is 5, and the radius is 3. The number of center nodes and periphery nodes, defined respectively as the nodes with eccentricity equal to radius and diameter, are both 36.
2) Degree assortativity, which measures how likely a node connects to a similar node, is -0.1517.
3) The clustering coefficient, which shows how well nodes in a graph tend to form cliques together, is 0.207. The transitivity, which considers the clustering of particular 3-node substructure, is 0.127. 
4) There are 60.748 unique cliques detected in the testnet. The modularity of the testnet, which measures the easiness of partitioning the graph into modules, is 0.0605.

As a baseline for comparison, we generate a random graph following the Erdos-Renyi~\cite{Erdos60onthe} (ER) model which generates an edge between each pair of nodes using the same probability, independently. It follows a binomial degree distribution and is commonly used as the network-analysis baselines. 
We use the same number of vertices and edges with the measured Ropsten network (that is, $n=588$ and $m=7496$) when generating the Erdos-Renyi random graph.
We run the graph generation algorithms for $10$ times and report the average properties of these random graphs in Table~\ref{tab:ropsten:comm}. 
{Particularly, the density is calculated by the number of measured intra-community edges divided by the number of total possible edges in that community. For instance, the density of a community of 92 nodes and 423 intra-community edges is $423/{8,000 \choose 2} = 0.10$.}
Besides, Table~\ref{tab:ropsten:comm} shows other two random graphs, namely configuration model~\cite{newman2003structure} (CM), and Barabasi-Alber~\cite{albert2002statistical} (BA). The former is generated using the same sequence of node degrees with the measured testnet, and the latter is generated using the same number of nodes ($n=588$) and same average node degree ($l'=26$).

Compared with the ER random graph, the measured Ropsten network has a much larger diameter, a smaller center size, a larger clustering coefficient, and more importantly, fewer cliques and lower modularity. This is similarly the case when comparing Ropsten with CM (except for CM's comparable diameter) and BA (except for BA's comparable number of cliques).
The implication is that a Ropsten network is much more resilient to network-partition attacks (e.g., eclipse and other DoS attacks) than these random graphs.

We also detect the communities in the Ropsten testnet, using the NetworkX tool~\cite{me:networkx:community} implementing the Louvain method described in~\cite{blondel2008fast}.
The results are in Table~\ref{tab:ropsten:comm}.
There are seven communities detected. 
The biggest community is community number two with $22\%$ of the nodes of the network. 
The average degree in the community is $19$, and $9\%$ of the nodes (i.e., 13 out of 142 nodes) only have a degree of 1.
By comparison, community number five contains $12.7\%$ of the nodes with the largest average degree $32.8$.

\ignore{
    \subfloat[Detected community structure]{%
\includegraphics[width=0.35\textwidth]{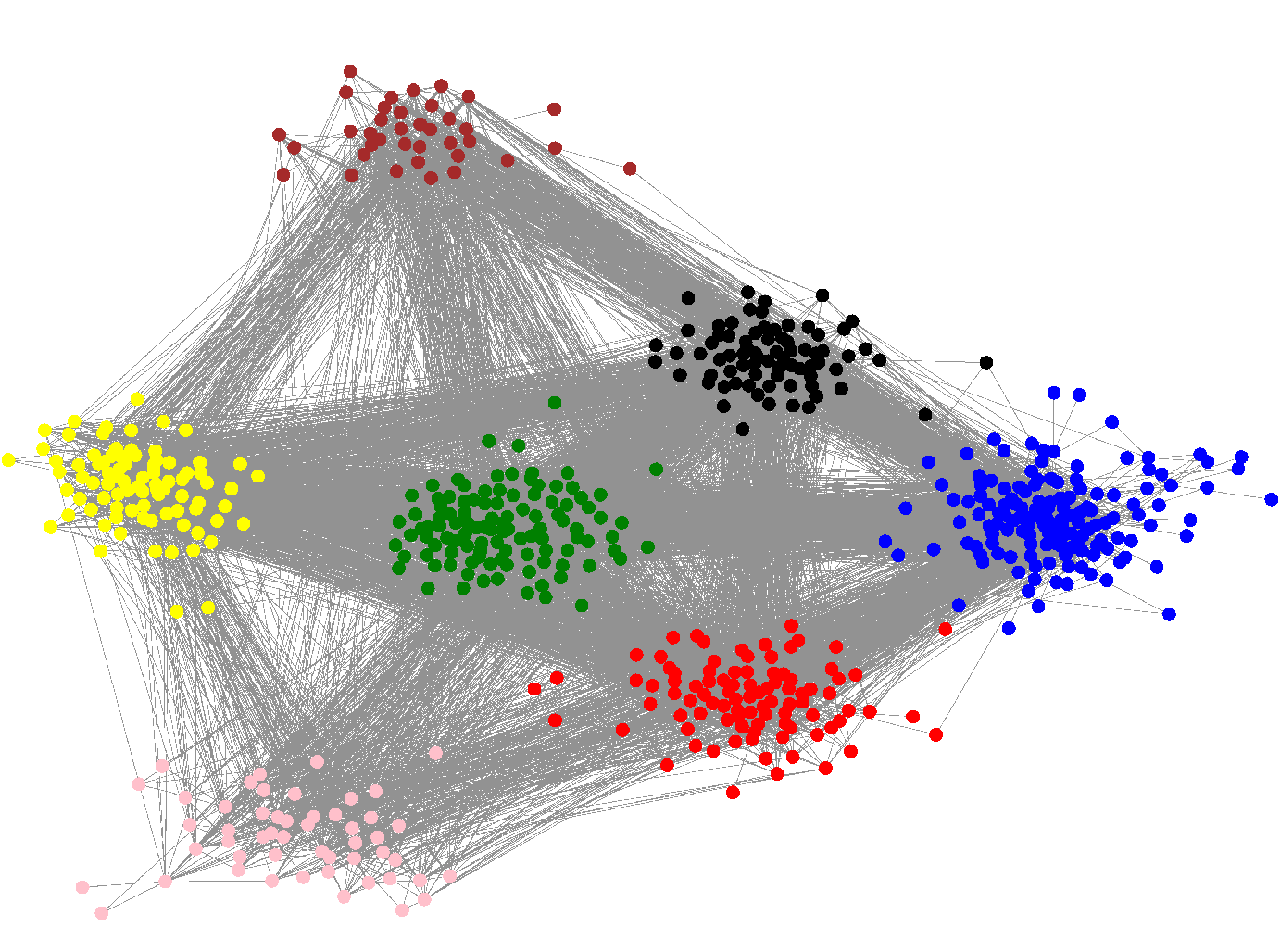}%
  \label{fig:ropsten}
    }%
    \subfloat[Geo distribution]{%
\includegraphics[width=0.35\textwidth]{figures_orig/ropsten_location.pdf}%
  \label{fig:ropsten:geo}
    }%

We plot the geographical distribution of the Ropsten testnet using GeoLite2 database service\footnote{\url{(https://github.com/maxmind/GeoIP2-python}}. Figure~\ref{fig:ropsten:geo} shows the Ropsten nodes in a world map and their connections. Most nodes are located in the United States, Europe, and East Asia. 

\begin{table}[!htbp] 
\caption{Nodes of large degree in Ropsten}
\label{tab:degree:large}\centering{
\begin{tabularx}{0.425\textwidth}{ |X|c|c|c| }
  \hline
Degree range & Count  & Degree range & Count \\ \hline
90-100 & 3  & 100-150 & 4  \\ \hline
150-200 & 6 & 418 & 1 \\ \hline
\end{tabularx}
}
\end{table}

}

\subsubsection{Summary of Rinkeby \& Goerli Results}

We conduct similar measurements on two other major Ethereum testnets, Rinkeby and Goerli. Here, we summarize what's noteworthy in the results while deferring the full presentation to Appendix~\ref{appdx:sec:rinkeby:goerli}.
From our measurement results, Rinkeby has smaller node degrees than Ropsten. Many Geth nodes in Rinkeby are with degrees smaller than $15$, and between degrees $15$ and $180$ the nodes are evenly distributed. In terms of graph statistics, Rinkeby's modularity (0.0106) is much lower than Goerli's (0.048) which is comparable with Ropsten's modularity (0.0605); this result implies that Rinkeby's the most resilient against network partitioning.

{{\bf Explaining the results}: In the measurement results, we consistently observe smaller modularity in testnets than that in random graphs. Full explainability of the measurement result is challenging and out of the scope of this paper. We take preliminary efforts to explain the measured results as follows.

We suspect the measurement results, particularly the discrepancy to the properties of random graphs and the much lower modularity, are due to the way Ethereum nodes choose/promote active links and the scale of the networks measured. Briefly, in the Ethereum protocol, a node maintains a ``buffer'' of inactive neighbors from which 50 active neighbors are selected in the case of existing active neighbors go offline. At the first glimpse, the presence of this buffer localizes the selection of active neighbors in a smaller candidate set than all the nodes as in the random graph, and it should facilitate forming the network of higher modularity. However, by looking more closely at the Ethereum protocol, a node $N$'s candidate buffer consists of node $N$'s inactive neighbors and node $N$'s inactive neighbors' inactive neighbors. For instance, with each node of $272$ inactive neighbors by default, the buffer size is $272*272=73984$ which is larger than the size of the testnets we measured. Thus, the effect of localization is not reflected in the testnet results. In fact, the deduplication of active neighbors (i.e., Ethereum clients, such as Geth, check if a recently promoted neighbor is already an active neighbor) may contribute to the much lower modularity in the measured testnets.
While here we explain the measurement results by qualitative analysis, we leave it to the future work the quantitative modeling and analysis of Ethereum network-connection protocols.
}

\subsection{Mainnet Measurement Results}
\label{sec:mainnet}

Measuring the mainnet's topology raises new challenges: 1) Due to ethical concerns, the measurement should not interfere the normal operation of live mainnet nodes. 2) Due to mainnet's large scale (about $8,000$ nodes and ${8,000 \choose 2}=\frac{1}{2}\cdot{}8,000\cdot{(8,000-1)}$ possible links) and the high price of Ether, measuring the entire network of mainnet incurs high cost, estimated to be more than $60$ million USD as will be analyzed.

To tackle the ethical challenge, we propose a \sysname extension to additionally verify certain conditions and ensure the non-interference to the service of target mainnet nodes. To bypass the high-cost challenge, in this work, we focus on measuring the topology of a small but critical subnetwork, instead of the entire mainnet.
{
 We conduct the measurement study on the mainnet on May 11th, 2021 and have spent $0.05858$ Ether (amount to $197.94$ USD at the price as of Aug 2021).}

{\bf Non-interference extension of \sysname}: Consider a measurement node $M$ runs \sysname against a target node $S$ in the Ethereum network $C$ ($S$ can be either $A$ or $B$ in our measurement primitive as in Figure~\ref{fig:example}). Suppose the measurement starts at time $t_1$ and ends at $t_2$. Node $M$ sets $tx_C$'s Gas price at $Y=Y_0$ and monitors the following two conditions. Only when both conditions are met, the measurement proceeds. 

\begin{itemize}
\item[V1)]
All blocks produced in $[t_1,t_2+e]$ are full in the sense that the Gas limit of each block is filled. $e$ denotes the expiration time of an unconfirmed transaction buffered in an Ethereum node. For instance, a Geth node would drop an unconfirmed transaction $e=3$ hours after it is submitted to the node, if it is not mined.
\item[V2)]
In the blocks produced during $[t_1,t_2+e]$, the included transactions have Gas prices higher than $Y_0$.
\end{itemize}

This extended \sysname achieves the following non-interference property: {\it The Ethereum blocks produced with the measurement turned on include the same set of transactions with the blocks produced with the measurement turned off.} We formally state the property and prove it in Appendix~\ref{appdx:sec:noninterference}.

{\bf Goal: mainnet's critical subnetwork}: With the above pricing strategy, measuring a pair of nodes on the mainnet costs $7.1\cdot{}10^{-4}$ Ether or $1.91$ USD (at the Ether price as of May, 2021). Thus, for the mainnet that consists of more than $8,000$ nodes, measuring all $\frac{1}{2}\cdot{}8,000\cdot{(8,000-1)}$ possible links would cost $22.845\cdot{}10^3$ Ether or more than $60$ million USD. We thus refrain from directly measuring the entire mainnet in this work. 

Instead, we choose a smaller but critical subnetwork of the mainnet to measure. Our observation is that in today's blockchain networks, essential transaction activities are centralized to few popular ``services'' that account for a small portion of the nodes in the network, such as popular transaction relay service (e.g., \infura\footnote{We anonymize the names of critical services tested to protect their service.} that relays $63\%$ of Ethereum transactions on the mainnet) and mining pools. 

We aim to answer the following research question: {\it RQ: Do Ethereum mainnet nodes prioritize the critical service nodes as their active neighbors?} 

To address the research, we design a measurement study on the mainnet that 1) discovers Ethereum nodes running behind the known popular services (including transaction relay and mining pools) and then 2) uses \sysname to measure the pair-wise connections among the discovered service-backend nodes. 

{\bf Step 1: Discovering critical nodes}: We discover the mainnet nodes on the backend of critical services. We use the approach described in the existing work~\cite{me:deter}. 
Briefly, the approach is to obtain the client version of the backend nodes by submitting the standard Ethereum RPC query (i.e., \texttt{web3\_clientVersion}) through the service frontend and to match the version against the ones in the Ethereum handshake messages received on a local ``supernode'' joining the mainnet, which is similar to the existing measurement work~\cite{DBLP:conf/imc/KimMMMMB18}. 

Using the above method, we discover the following mining-pool nodes on the mainnet: $59$ \sparkpool nodes, $8$ \twominers nodes, $6$ \icanminingru nodes, $2$ \digipools nodes, $2$ \cheapethpool nodes, and $1$ \ubiqpool node.
We also discover the following transaction-relay nodes on the mainnet: $48$ \infura nodes and $1$ \blockdaemon node.
When discovering the nodes, we use the codename revealed through the \texttt{web3\_clientVersion} query. The measurement result is consistent with~\cite{me:deter}.

\begin{table}[!htbp] 
\caption{Connections among critical nodes}
\label{tab:critical:connect}\centering{\small
\begin{tabularx}{0.35\textwidth}{ |l|l|l|X| }
  \hline
Type  & Conn. & Type & Conn. \\ \hline
\infura - \sparkpool   & \cmark & \sparkpool - \sparkpool     & \xmark \\ \hline
\infura - \twominers     & \cmark & \sparkpool - \twominers       & \cmark \\ \hline
\infura - \icanminingru  & \cmark & \sparkpool - \digipools      & \cmark \\ \hline
\infura - \digipools    & \cmark & \sparkpool - \icanminingru    & \cmark \\ \hline
\blockdaemon - \sparkpool  & \xmark & \twominers - \twominers     & \cmark \\ \hline
\blockdaemon - \twominers    & \xmark & \twominers - \icanminingru  & \cmark \\ \hline
\blockdaemon - \icanminingru & \xmark & \twominers - \digipools    & \cmark \\ \hline
\blockdaemon - \digipools   & \xmark & \icanminingru - \digipools & \cmark \\ \hline
\blockdaemon - \infura     & \xmark & \infura - \infura    & \cmark \\ \hline
\end{tabularx}
}
\end{table}

{\bf Step 2: Measuring topology among critical nodes}: We run the extended \sysname to detect whether critical nodes discovered as above are connected with each other. We consider three possible connection types: the inter-connection between a mining-pool node and a relay-service node, the connection between two mining-pool nodes and the connection between two relay-service node. 
For each case, we select random nodes from each service and measure all possible links. 
For measuring the connection between ``\infura - \sparkpool'', for instance, we select two random \infura nodes and two random \sparkpool nodes, and measure the four combinations of links. 
In addition, we select two nodes for \twominers and select one node for each of the services: \blockdaemon, \icanminingru, \digipools. In total, we choose 9 mainnet nodes.

We report the result in Table~\ref{tab:critical:connect}. We make the following observation: 1) A node behind relay service \infura connects to all tested mining pools and other \infura nodes. It does not connect to other relay services such as \blockdaemon. 
2) The single node behind relay service \blockdaemon does not connect to any mining pools or other relay service. Here, \blockdaemon's node may randomly choose neighbors as vanilla Ethereum clients do. 3) Nodes behind all mining pools connect to nodes of the same pool and other pools. They also connect to \infura. The only exception is that \sparkpool nodes do not connect to other \sparkpool node. 

{{\bf Explaining the results}: There are two possible explanations of the results: a) \infura and all mining pools run supernodes internally, which connect to all other nodes. \blockdaemon runs a regular node that declines incoming connection requests once its active neighbors are full. 
b) An \infura node prioritizes the connections to other \infura nodes and mining-pool nodes. It does not prioritize connecting to other RPC-service nodes like \blockdaemon. So are the mining pool nodes.}
\vspace{0.15in}

{\subsection{Summary of Measurement Costs/Time} 
We summarize the measurement costs/time in Table~\ref{tab:measure:summary}, which reports the actual Ether cost spent for measuring the testnets and the estimated cost of measuring the full topology of mainnet. The mainnet cost is estimated by multiplying the pairwise-measurement cost by the number of possible edges in the network (as mentioned before). Note that in the mainnet, the measurement transactions' Gas prices are set to be higher than at least $10\%$ of the pending transactions in the \texttt{mempool} (for estimation purposes, we assume the target node's \texttt{mempool} has the same content as the measurement node's mempool). 
}

{\begin{table}[!htbp] 
\caption{Summary of measurement studies on the testnets/mainnet. \# refers to ``number''.}
\label{tab:measure:summary}\centering{\small
\begin{tabularx}{0.475\textwidth}{ |X|X|c|c|X| }
  \hline
Network & Size (\# of nodes) & Cost (Ether) & Date & Duration (hours) \\ \hline
Ropsten & 588 & 0.067 & Oct. 30, 2020 & 12\\ \hline
Rinkeby & 446 & 2.10 & Nov. 15, 2020 & 10\\ \hline
Goerli & 1025 & 0.62 & Oct. 20, 2020 & 20\\ \hline
mainnet & 9 & 0.05858 & May. 15, 2021 & 0.5\\ \hline
\end{tabularx}
}
\end{table}
}

\section{Ethical Discussion}
\label{sec:discuss}

In this work, we use \sysname to measure testnets. While the approach is active measurement (to refill underwhelmed \texttt{mempool} in the testnet), the testnets do not run business, and the possible service interruption to the testnets will have limited impacts.
We also measure a limited sub-network on the Ethereum mainnet. 
As analyzed before in \S~\ref{sec:mainnet}, the presence of measurement using the \sysname extension does not affect what set of transactions are included in the blockchain.
A more formal statement is in Theorem~\ref{thm:noninter} which is proven in Appendix~\ref{sec:noninterfere:analysis}. 
We believe \sysname's impact to normal transactions when measuring the mainnet is small.

\section{Conclusion}
\label{sec:conclude}

This work presents \sysname, a measurement study that uncovers Ethereum's network topology by exploiting transaction replacement and eviction policies. \sysname achieves the perfect precision and high recall. A parallel schedule is proposed to apply the pair-wise measurement to large-scale networks.
\sysname uncovers the topology of three major Ethereum testnets, which show their difference to random graphs and high resilience to network partitioning. We also use \sysname to measure critical service interconnection in the mainnet which reveals biased neighbor selection strategies by top mining pools and relay service nodes.

%% file: text/wellformatted_appendix.tex
\noindent{}{\LARGE \bf Appendices}

\section{TxProbe's Applicability to Ethereum: Additional Details}
\label{appdx:sec:txprobe:applicable} 

TxProbe is inapplicable to measuring the topology of Ethereum network, due to two distinct features in Ethereum: 1) Ethereum's propagation model where transactions can be directly propagated without announcement, and 2) Ethereum's account-based model where the balance state can be arbitrarily recharged while a Bitcoin balance can only transit one way, from unspent to spent (as in its UTXO model). In the main text, we explain why TxProbe's inapplicable in Ethereum due to 2). Here, we explain TxProbe's inapplicable in Ethereum due to 1).

Briefly, before sending $tx_A$, TxProbe sends two double-spending transactions respectively to node $A$ and $B$ such that $tx_A$ on node $B$ will become an orphan transaction that is not propagated further. However, with Ethereum's account model, $tx_A$ on node $B$ would not necessarily be an orphan transaction (or equivalently a future transaction in Ethereum's jargon). It can be an overdraft transaction that is propagated. More detailed description is below:
 
TxProbe~\cite{DBLP:conf/fc/Delgado-SeguraB19} actively measures Bitcoin network topology, by exploiting its handling of orphan/double-spending transactions in transaction propagation. Other works~\cite{DBLP:conf/fc/GrundmannNH18} measure the Bitcoin topology using a similar approach. In the following, we describe the working of TxProbe~\cite{DBLP:conf/fc/Delgado-SeguraB19} in detail, with the purpose to discuss its applicability to measuring Ethereum topology.
Suppose using TxProbe to measure the connection between nodes $A$ and $B$. The measurement node first sends to nodes $A$ and $B$ two double-spending transactions, say $tx_A'$ and $tx_B'$. It then sends the third transaction $tx_A$ spending $tx_A'$ to node $A$. It observes the presence of the transaction $tx_A$ on node $B$. If it is present, there is a connection between nodes $A$ and $B$. 

Applying TxProbe to measuring Ethereum topology is unfeasible due to Ethereum's account model: Ethereum adopts a different model to store ledger states than Bitcoin, and the definition of orphan transactions in Ethereum is different than that in Bitcoin. Specifically, Ethereum stores ledger state (how much cryptocurrency an address/account has) in per-account balances, namely the account model, while Bitcoin stores the balance state in per-transaction UTXO, namely the UTXO model. Under UTXO, an orphan transaction is a transaction that spends an input whose state is yet to be determined. This makes the third transaction in the TxProbe protocol an orphan on Node $B$, which does not propagate. However, under the account model, an orphan transaction (or so-called future transaction) is one with a noncontinuous nonce to any previous transaction, where a nonce is per a sender account. Thus, the third transaction that spends a double-spending transaction is not necessarily an orphan, as it may be an over-drafting transaction with a valid nonce (in which case, the transaction will be propagated by node $B$).

\section{Local Validations}
\label{sec:validate:local}

\subsection{Local Validation of Serial Measurement}

{\bf Local evaluation}: In this experiment, we set up three local Ethereum nodes, mutually connected and without the communication to any external Ethereum nodes. The three nodes represent nodes $M$, $A$ and $B$ in the \sysname protocol. In this local environment, we aim at evaluating \sysname's correctness with respect to varying \texttt{mempool} sizes on node $A$. 

On node $A$, we vary node $A$'s \texttt{mempool} sizes from $3120$ to $9120$. In the experiments, we also populate the three-node network with a varying number of pending transactions $tx_O$'s, such as 1, 1000, 2000, and 3000 such transactions. 

\begin{figure}[!ht]
  \begin{center}
  \includegraphics[width=0.5\textwidth]{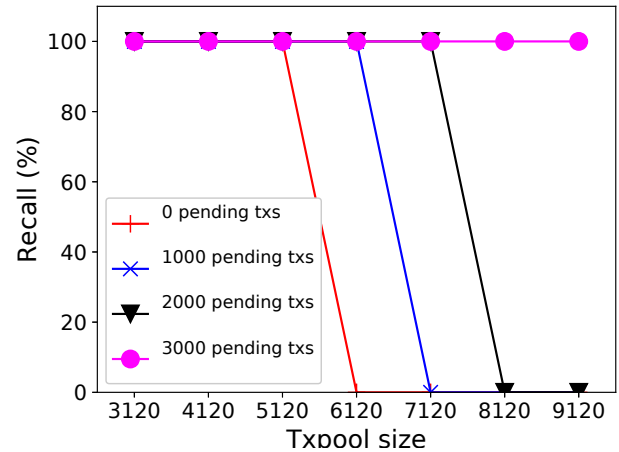}%
  \end{center}
  \caption{Recall with increasing \texttt{mempool} size}
  \label{fig:recall:mempool}
\end{figure}

We use the measurement results from \sysname and compare it against the ground-truth to report the recall in Figure~\ref{fig:recall:mempool}. The result shows that given \texttt{mempool} size $X$ and the number of pending transactions $X'$, \sysname achieves $100\%$ recall when $X-X'<=5120$. Otherwise, the recall drops to $0\%$.

\ignore{
\begin{figure*}[!bpht]
  \begin{center}
    \subfloat[Community in Rinkeby]{%
\includegraphics[width=0.25\textwidth]{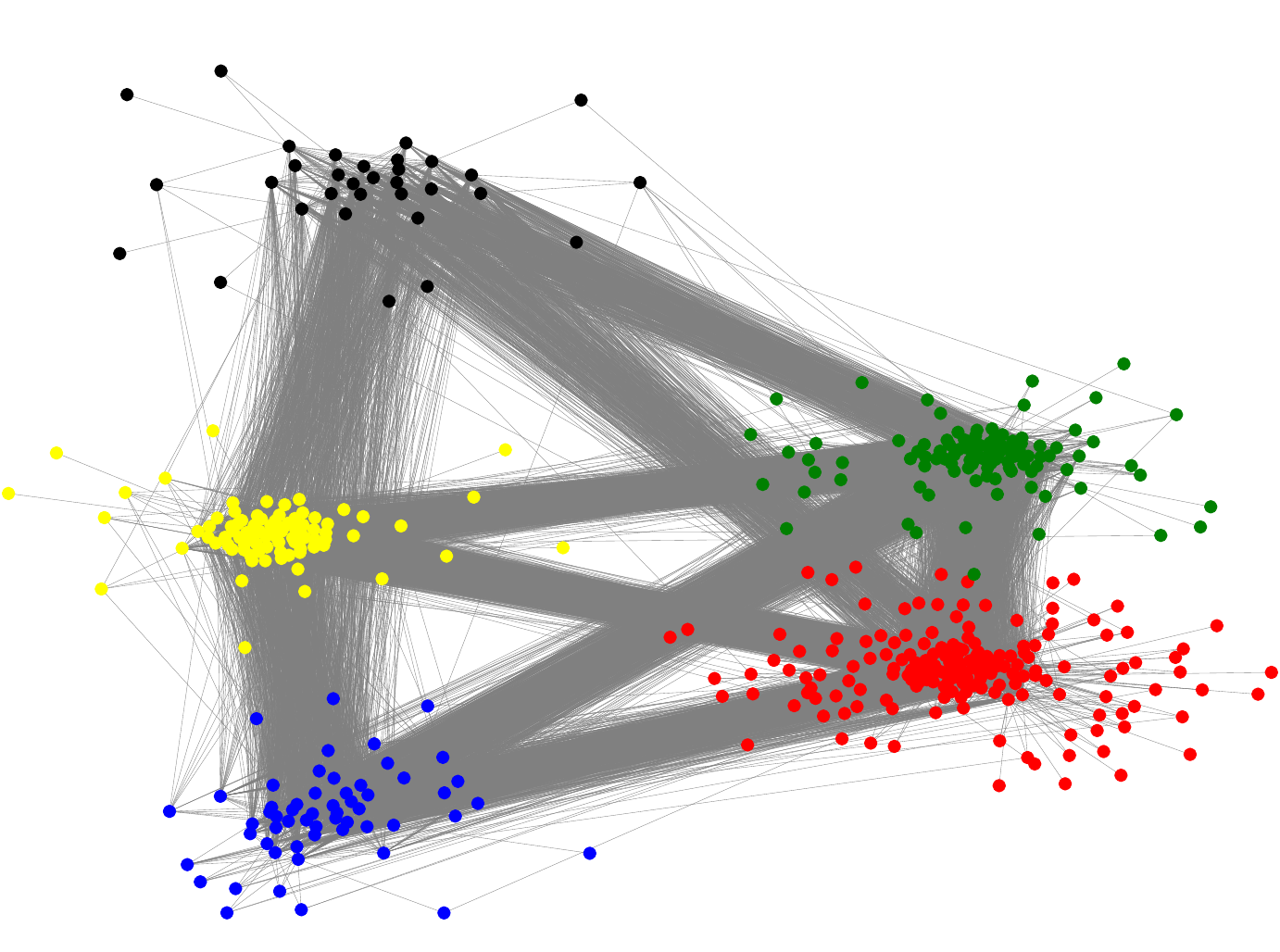}
  \label{fig:rinkeby}
    }%
    \subfloat[Geo distribution of Rinkeby]{%
    \includegraphics[width=0.25\textwidth]{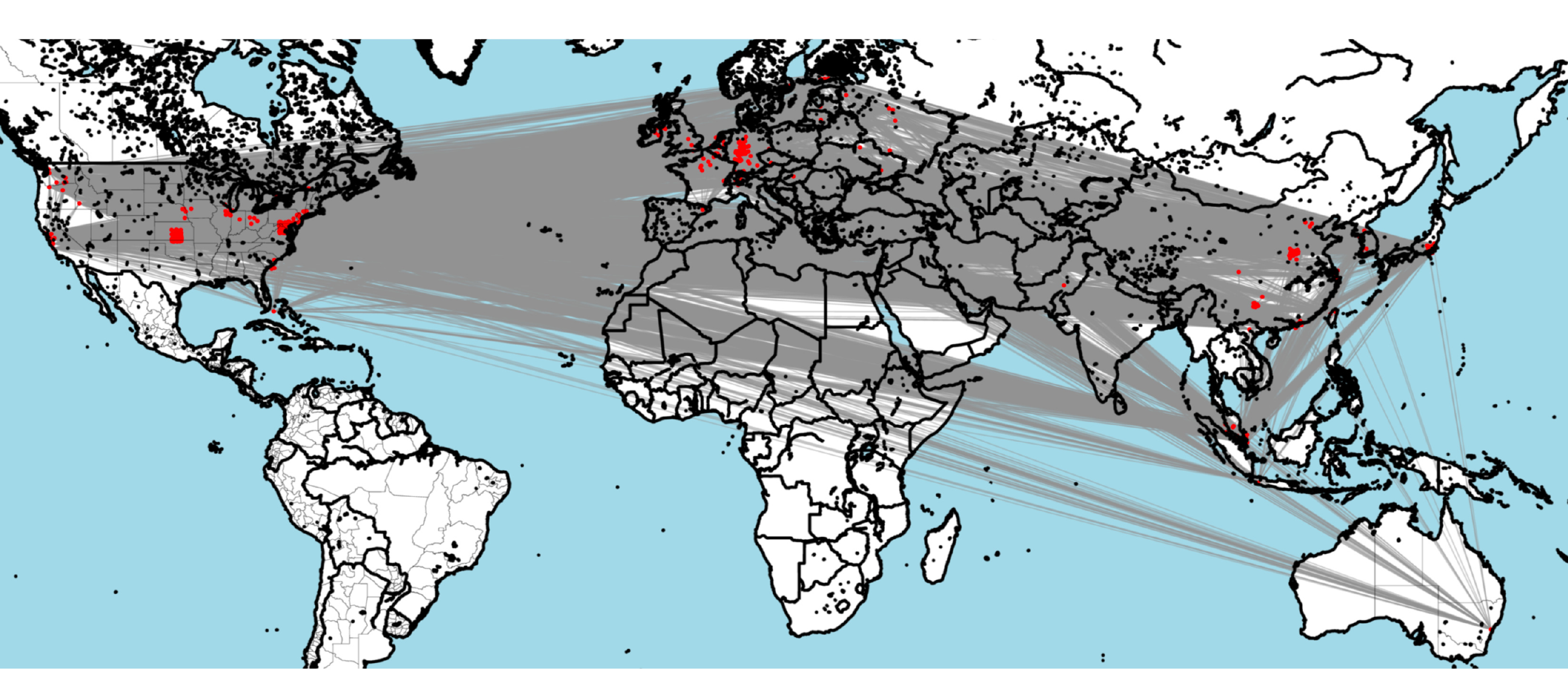}%
  \label{fig:rinkeby:geo}
    }%
    \subfloat[Community in Goerli]{%
\includegraphics[width=0.25\textwidth]{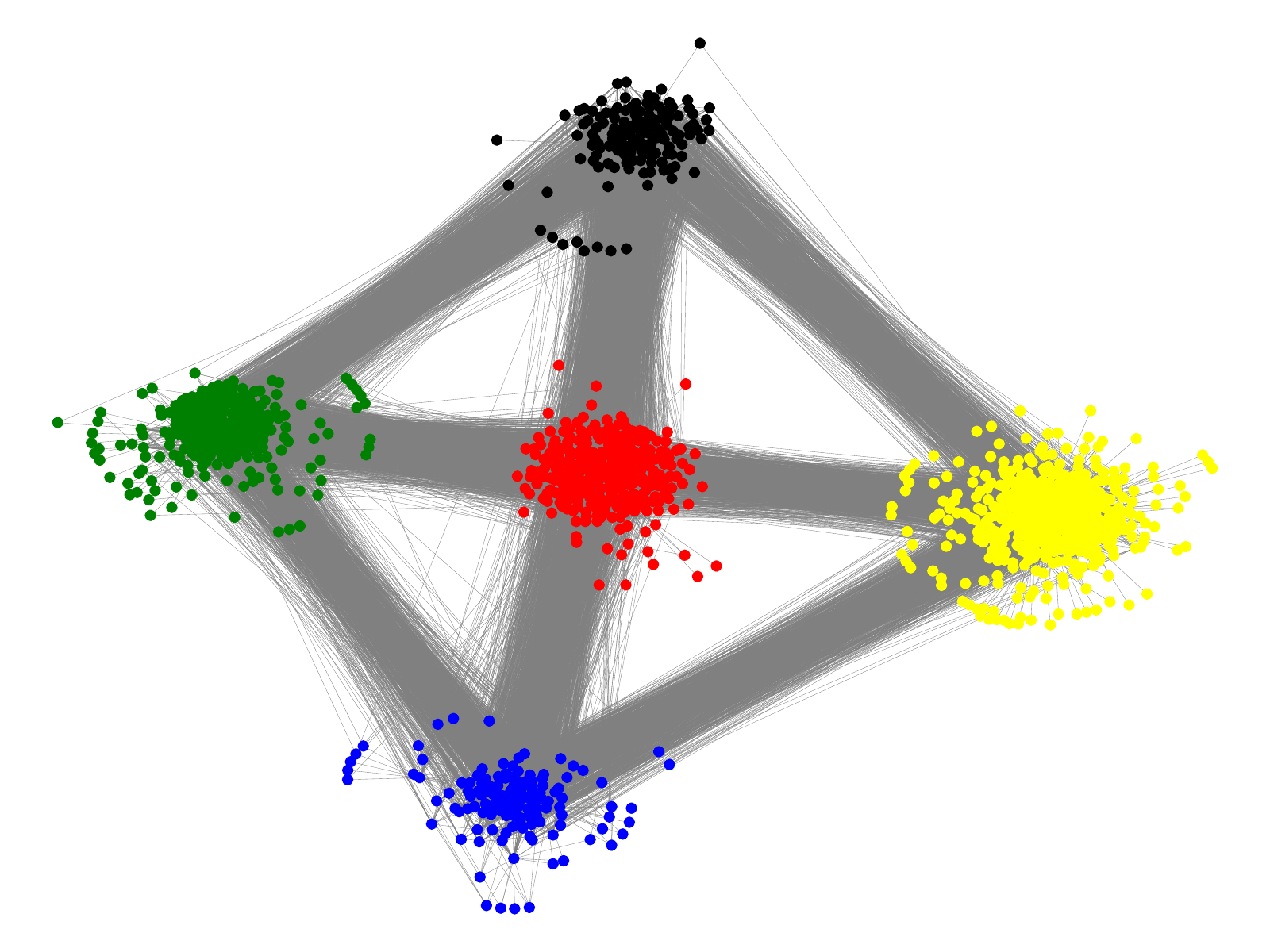}%
  \label{fig:goerli}
    }%
    \subfloat[Geo distribution of Goerli]{%
    \includegraphics[width=0.25\textwidth]{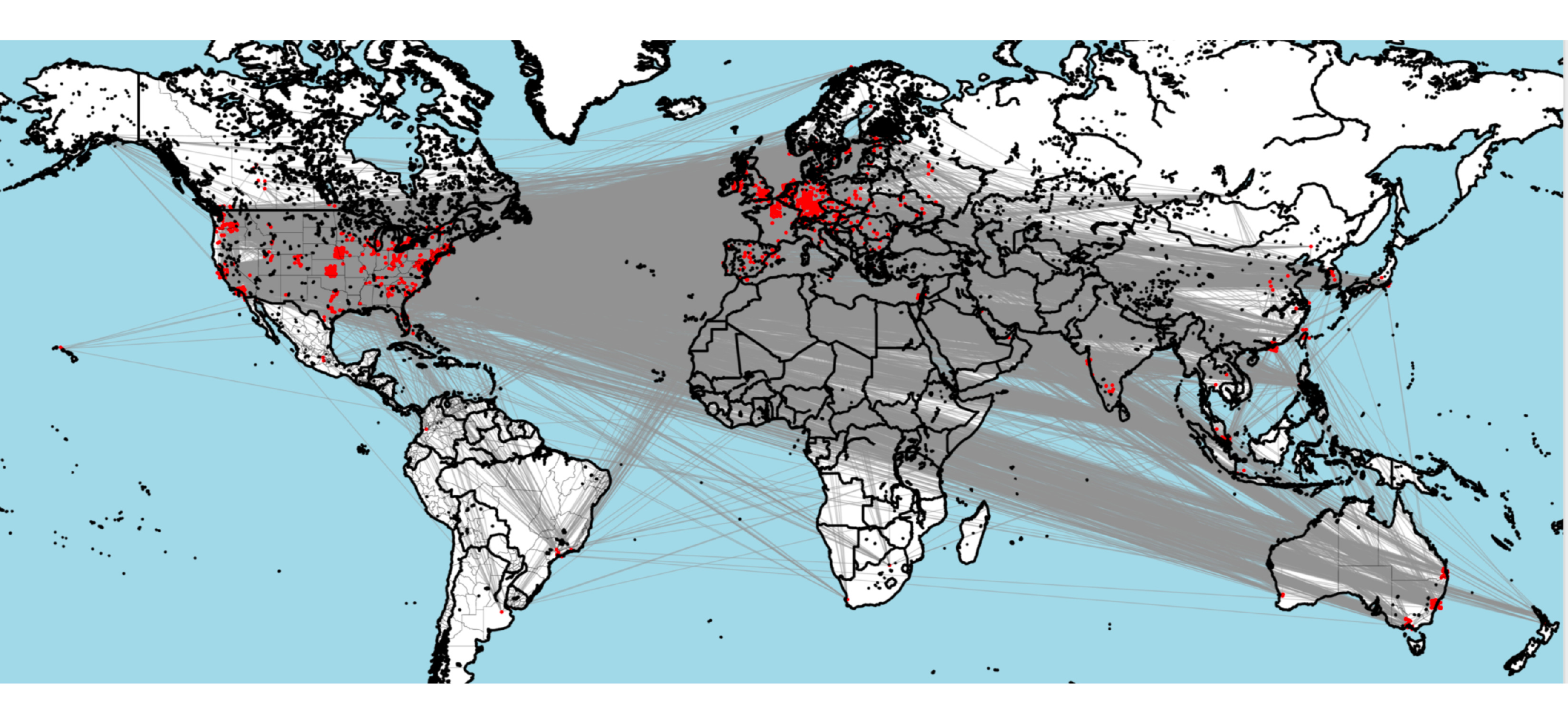}%
  \label{fig:goerli:geo}
    }%
  \end{center}
  \caption{Visualization of measured Rinkeby and Goerli networks}
\end{figure*}
}

This validation study implies that matching the number of pending transactions to the actual \texttt{mempool} size is crucial to achieving $100\%$ measurement recall.

\subsubsection{Local Validation of Parallel Measurement}

We conduct a validation of the parallel measurement method in a local environment without connection to a remote Ethereum network. Here, the measurement node $M$, two nodes $A_1$ and $A_2$, as well as node $B$, are run on local machines under our control.

In terms of the connections among $A_1$, $A_2$ and $B$, there are two permutations of three possibilities, that is, $P(3,2)=8$. Given the symmetry (e.g., $\langle{}A_2,B\rangle{}$ is equivalent to $\langle{}A_1,B\rangle{}$), we consider six possibilities as listed below. For each possibility, we use \sysname to conduct measurements for $100$ times. The final result is positive if any of the $100$ measurements returns a positive result (i.e., there is a connection). 
Then by comparing the measurement results and the ground-truth, we obtain the precision and recall of the measurement results.

\begin{table}[!htbp] 
\caption{Recall and precision of using \sysname on local nodes}
\label{tab:perm}\centering{\small
\begin{tabularx}{0.375\textwidth}{ |X|c|c| }
\hline
& Recall & Precision \\ \hline
$\langle{}A_1,A_2\rangle{}$, $\langle{}A_1,B\rangle{}$, $\langle{}A_2,B\rangle{}$ & 100\% & 100\% \\ \hline
$\langle{}A_1,A_2\rangle{}$, $\langle{}A_1,B\rangle{}$ & \multirow{2}{*}{100\%} & \multirow{2}{*}{100\%} \\
$\langle{}A_1,A_2\rangle{}$, $\langle{}A_2,B\rangle{}$ & & \\ \hline
$\langle{}A_1,A_2\rangle{}$ & 100\% & 100\% \\ \hline
$\langle{}A_1,B\rangle{}$, $\langle{}A_2,B\rangle{}$ & 100\% & 100\% \\ \hline
$\langle{}A_1,B\rangle{}$ (also $\langle{}A_2,B\rangle{}$) & 100\% & 100\% \\ \hline
Null & 100\% & 100\%
\\\hline
\end{tabularx}
}\end{table}

As can be seen from table~\ref{tab:perm}, all results are with $100\%$ recall and precision, even when there is a connection between $A_1$ and $A_2$. The theoretic measurement inaccuracy when $A_1$ connects to $A_2$ seems unlikely to occur in practice.

\section{Measurement Extension for Non-interference}
\label{appdx:sec:noninterference}

\subsection{Extending \sysname with Non-interference Verification}

{\bf Goals}: When running our \sysname against an operational Ethereum network, notably the mainnet, it is required that the measurement does not interfere with the normal operations of the network; for instance, the \sysname should not evict any transactions that are otherwise included in the blockchain. This non-interference property is formally described in our analysis framework in \S~\ref{sec:noninterfere:analysis}. 

{\bf Design rationale}: To design \sysname extension for assurance of non-interference, one possible approach is to set a low Gas price (i.e., \sysname's parameter $Y$) and prove the non-interference, {\it a priori}, by considering the theoretically worst case that could occur after the measurement starts. Our initial design follows this approach, but just to find it is unfeasible with the current Ethereum-node settings.\footnote{To be specific, Geth's default \texttt{mempool} length $5120$ is too small to feed all the blocks produced in the three-hour span (expiration time), by considering the worst case that no new transactions are submitted to the Ethereum network after the measurement.}

Instead of proving non-interference a priori, we aim at verifying the non-interference a posteriori. That is, the measurement node $M$ initially sets a conservatively low Gas price (e.g., based on heuristics) and conducts the measurement. Meanwhile the node monitors several conditions on the tested network during and after the measurement, in order to establish non-interference {\it posterior}.

{\bf Measurement extension for verification}: Consider a measurement node $M$ runs \sysname against a subject node $S$ in the Ethereum network $C$ ($S$ can be either Node $A$ or $B$ as in our serial-measurement model in Figure~\ref{fig:example}). The measurement starts at time $t_1$ and ends at $t_2$. Node $M$ sets a low Gas price at $Y=Y_0$ and monitors the blockchain on the following conditions:

\begin{itemize}
\item[V1)] 
All blocks produced in $[t_1,t_2+e]$ are full in the sense that the Gas limit of each block is filled. Here, $e$ denotes the expiration time of a transaction in Ethereum-node \texttt{mempool}, for instance, $e=3$ hours in Geth by default.
\item[V2)] 
In the blocks produced in $[t_1,t_2+e]$, all transactions' Gas prices are higher than the preset Gas price $Y_0$.
\end{itemize}

\subsection{Non-interference Analysis}
\label{sec:noninterfere:analysis}

In this subsection, we first define what the measurement interference means. We then prove that verified Conditions V1 and V2 ensure non-interference on the measured nodes.

Intuitively, non-interference means the action of measurement does not affect what blocks are produced by the Ethereum network being tested. In other words, with and without the measurement $P$, the blocks produced by the Ethereum network should be the same.

Formally, we consider a node $M$ runs a measurement process against a subject node $S$, which is connected to the rest of an Ethereum network $C$. In the case that \sysname is used to detect the link between $A$ and $B$, $S$ can be either node $A$ or $B$. The measurement process starts at time $t_1$ and ends at time $t_2$. 

\begin{definition}
Consider a measurement process parameterized by $P(M,S,C,t_1,t_2)$. Denote by $\{b_i\}$ the sequence of blocks produced by the Ethereum network $(S,C)$ in period $[t_1,t_2+e]$.

Now consider a hypothetical world in which the measurement did not occur at $t_1$ and the Ethereum network produces the sequence of block headers with $\{b'_i\}$. The hypothetical world is deterministic in the sense that it produces the same block from the same miner at the same time with the actual world with measurement, that is, block $b'_i$ has the same timing with $b_i$.

$P(M,S,C,t_1,t_2)$ does not interfere with the measured Ethereum network $(S,C)$, if and only if the transactions included in each block $b_i$ (i.e., the block at the index $i$ of the sequence) in the actual world with measurement are identical to those included in block $b'_i$ in the hypothetical world without measurement.
\end{definition}

\begin{theorem}
Consider a measurement $P(M,S,C, t_1,t_2)$ is conducted using the method of \sysname. If Conditions V1 and V2 hold, $P$ does not interfere with the Ethereum network $(S,C)$.
\label{thm:noninter}
\end{theorem}
\begin{proof}

Generally speaking, blocks can be produced by node $S$ or other nodes in Ethereum network $C$. Because \sysname will not evict transactions on nodes besides $S$, the measurement will not affect the block produced by nodes other than $S$. Thus, we consider in this proof the ``worst case'' that all blocks $\{b_i\}$ are produced by node $S$.

Due to the design of \sysname, the measurement process will evict only the transactions in $S$'s \texttt{mempool} (as late as of time $t_2$) whose Gas prices are lower than $Y_0$. For other transactions whose Gas prices are higher than $Y_0$ and transactions submitted after $t_2$, measurement process $P$ will not affect them.

Now, we are ready to prove the theorem by contradiction: Assuming there is interference under Conditions V1 and V2, our goal of the proof is to find contradictory. That is, with V1 and V2, there is at least one transaction included in a block produced hypothetically without measurement, say $b'_i$, and that is not included in the corresponding actual block $b_i$. We name this transaction by $tx_l$. Since the measurement will only affect the transactions with Gas prices lower than $Y_0$, $tx_l$'s Gas price must be lower than $Y_0$. 

Because of $V1$, $b_i$ must be full. Thus, there must be a ``victim'' transaction in the hypothetical world, say $tx_h$, that is included in $b_i$ but is replaced by $tx_l$ in $b_i'$. Because $tx_h$'s Gas price is higher than $Y_0$ and is not affected by the presence of measurement, $tx_h$ must reside in the node's \texttt{mempool} in the hypothetical world without measurement.

Now, we can constitute a scenario in the hypothetical world that Miner $S$ is faced with two transactions in its \texttt{mempool}, $tx_l$ and $tx_h$. To make $tx_l$ in the blockchain, $S$ must prioritize $tx_l$ (with Gas price lower than $Y_0$) over $tx_h$ (with Gas price higher than $Y_0$) to mine. This {\it contradicts} with the property in Ethereum client implementations (both Geth and Parity) that transactions of higher Gas price have higher priority to be mined than those of lower Gas price. Note that here both $tx_l$ and $tx_h$ have small enough Gas and can fit into block $b_i$ under the block Gas limit.
\end{proof}
            
\section{Measurement Results of Rinkeby \& Goerli}
\label{appdx:sec:rinkeby:goerli}

\begin{figure*}
\begin{minipage}{.3\textwidth}
\includegraphics[width=0.975\textwidth]{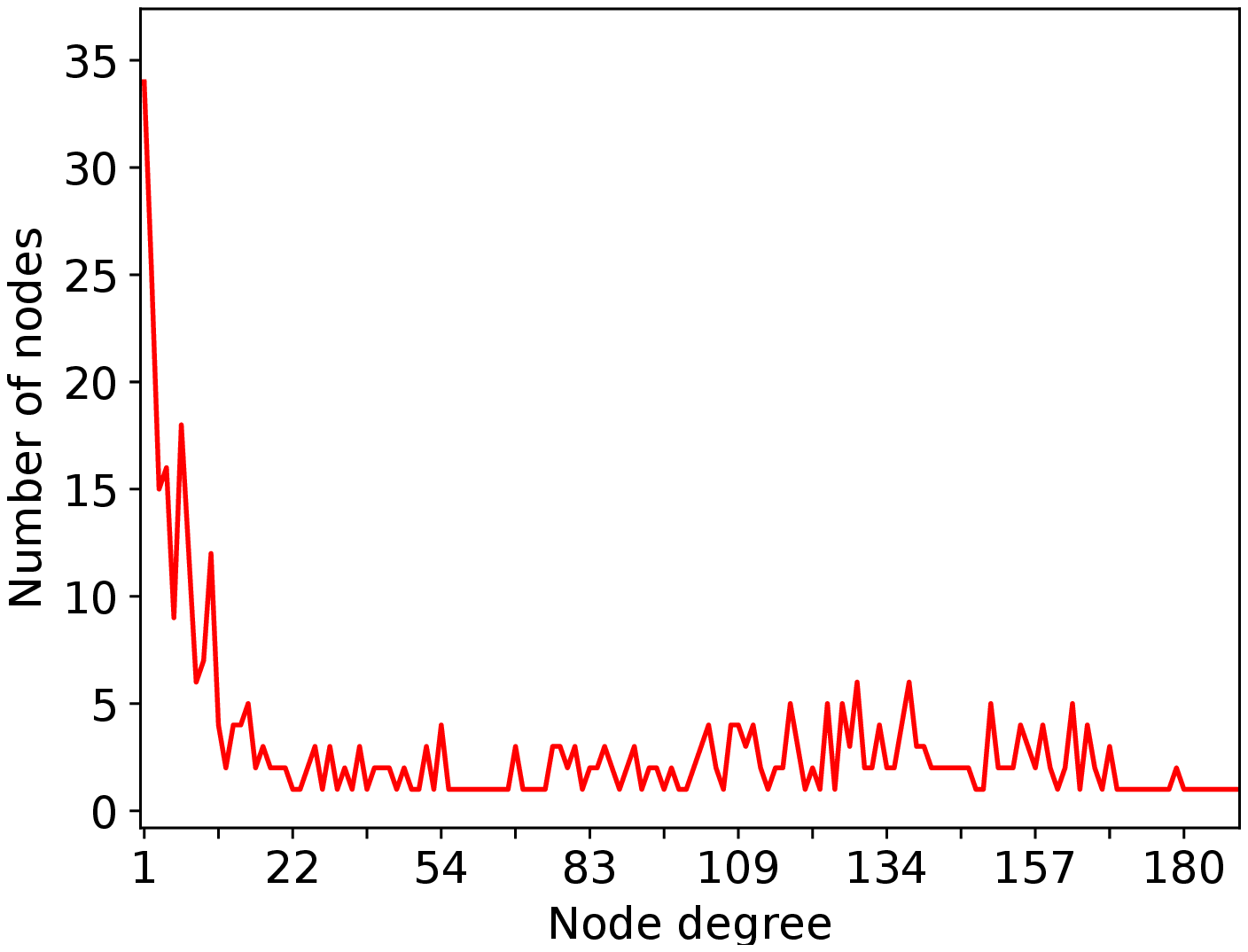}
  \caption{Degree distribution in Rinkeby}
  \label{fig:rinkeby:degree}
\end{minipage}
\begin{minipage}{.3\textwidth}
\includegraphics[width=\textwidth]{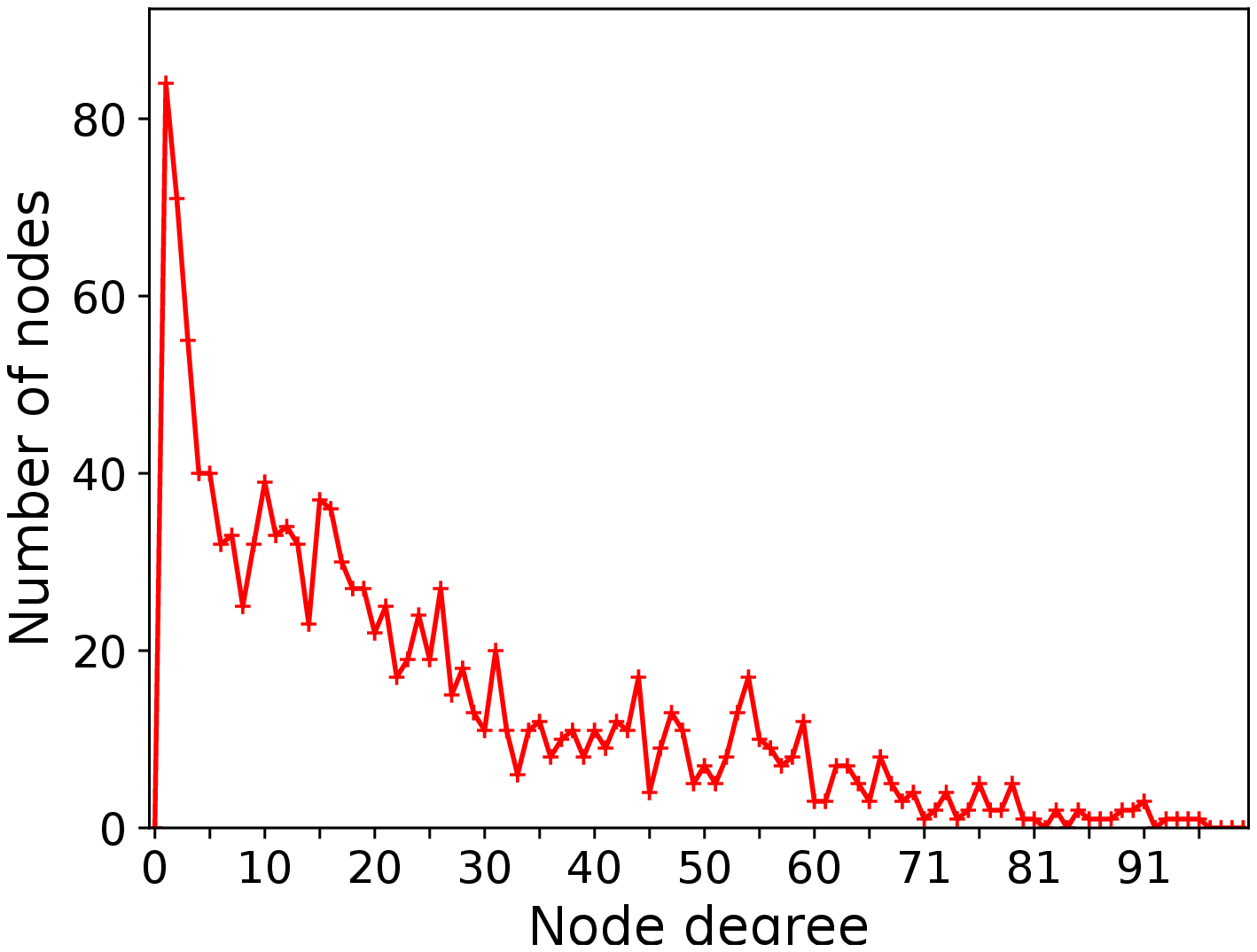}
  \caption{Node degree distribution in Goerli}
  \label{fig:goerli:degree}
\end{minipage}
\begin{minipage}{0.375\textwidth}
\caption{Nodes of large degree in Goerli}
\label{tab:goerli:degree:large}\centering{\small
\begin{tabularx}{\textwidth}{ |X|c|c|c| }
  \hline
Degree range & Count  & Degree range & Count \\ \hline
100-150 & 12 & 150-200 & 3 \\ \hline
200-300 & 4 & 300-500 & 3 \\ \hline
697   &  1 & 711 & 1 \\ \hline
\end{tabularx}
}
\end{minipage}
\end{figure*}

\begin{table}[!htbp] 
\caption{Graph properties of the Rinkeby testnet}
\label{tab:graphproperties:rinkeby}\centering{\footnotesize
\begin{tabularx}{0.45\textwidth}{ |X|X|X|X|X| }
  \hline
&
Measured Rinkeby & ER (n=446, m=15380) & CM & BA (n=446, l=69)
\\ \hline
Diameter & 4 & 2.7 & 4.6 & 2.0 
\\ \hline
Periphery size & 203 & 512.0 & 76.1 & 446.0
\\ \hline
Radius & 3 & 2.0 & 3.0 & 2.0 
\\ \hline
Center size & 243 & 442.4 & 233.3 & 446.0 \\ \hline
Eccentricity & 3.455 & 2.008 & 3.4953 & 2.0 
\\ \hline
Clustering coefficient & 0.4375 & 0.1548 & 0.3407 & 0.3592  \\ \hline
Transitivity & 0.4981 & 0.1548 & 0.3589 & 0.3513
\\ \hline
Degree assortativity & -0.03202 & -0.001536 & -0.03275 & -0.04555 \\ \hline
Clique number & 274775.0 & 150.6 & 383.2 & 82.5 \\ \hline
Modularity & 0.01063 & 0.08198 & 0.07332 & 0.05310 \\ \hline
\end{tabularx}
}
\end{table}

\begin{table}[!htbp] 
\caption{Graph properties of the Goerli testnet}
\label{tab:graphproperties:goerli}\centering{\footnotesize
\begin{tabularx}{0.45\textwidth}{ |X|X|X|X|X| }
  \hline
&
Measured Goerli & ER (n=1025, m=18530) & CM & BA (n=1025, l=36)
\\ \hline
Diameter & 5 & 3.0 & 5.1 & 3.0
\\ \hline
Periphery size & 23 & 1025.0 & 31.3 & 866.3 
\\ \hline
Radius & 3 & 3.0 & 3.0 & 2.0
\\ \hline
Center size & 115 & 1025.0 & 154.7 & 158.7   
\\ \hline
Eccentricity & 3.775 & 3.0 & 3.911 & 2.845 
\\ \hline
Clustering coefficient & 0.0354 & 0.0355 & 0.1281 & 0.1380   
\\ \hline
Transitivity & 0.09616 & 0.0354 & 0.1052 & 0.1374
\\ \hline
Degree assortativity & -0.1573 &  -0.0036 & -0.0742 & -0.0050
\\ \hline
Clique number & 134.49 & 416.8 & 1007.2 & 63.4
\\ \hline
Modularity & 0.048 & 0.132 & 0.125 & 0.084 \\ \hline
\end{tabularx}
}
\end{table}



We similarly apply the \sysname method to measure the Rinkeby testnet. Compared with Ropsten, Rinkeby is more heavily used and the \texttt{mempool}s there contain more transactions. For instance, on our local node $M$ connected to Rinkeby, it is not uncommon that the \texttt{mempool} has more than $4500$ transactions. We thus estimate the median Gas price in the \texttt{mempool} (using the method described in \S~\ref{sec:gasprice}) and use it as $tx_C$'s Gas price. 

Noteworthy is that during this measurement, we found when our measurement node $M$ sends future transactions (as in Step \ballnumber{\scriptsize 2}) to certain nodes in Rinkeby, these nodes return the same future transactions back to node $M$. To avoid overloading $M$ with the future transactions bounced back, we modify the Geth client running on $M$ to discard figure transactions received from other nodes.

We present the similar measurement metrics of Rinkeby with Ropsten. The node degree distribution is in Figure~\ref{fig:rinkeby:degree} where node degrees are distributed from $1$ to $180$. There are many nodes with degree smaller than $15$, and between degrees $15$ and $180$ the nodes are evenly distributed. Graph statistics of Rinkeby, in comparison with the three random graphs, are presented in Table~\ref{tab:graphproperties:rinkeby}, where the measured testnet similarly shows most traits, such as with much lower modularity, which implies the testnet's higher resilience to network partitioning. Particularly, there are many more cliques found on Rinkeby than on the random graphs, which corroborates the low modularity of the testnet and hardness to partition its topology.

Compared with Ropsten, Rinkeby has a much larger center size (more nodes in the center of the graph), a higher transitivity (more likely the adjacent nodes are connected) and a lower-level modularity (harder to partition the graph into densely connected modules).

We conducted a similar measurement study on Goerli, another Ethereum testnet, and present results in node degree distribution in Figure~\ref{fig:goerli:degree} and Figure~\ref{tab:goerli:degree:large}, and graph statistical properties in Table~\ref{tab:graphproperties:goerli}. Notably, there are nodes in the Goerli network that are globally connected and are with very high degrees (e.g., more than $700$ neighbors). It has a very low clustering coefficient ($0.0354$) compared with those of Rinkeby ($0.4375$) and Ropsten ($0.207$). In terms of modularity, Goerli ($0.048$) is comparably lower than Ropsten ($0.0605$), and is much higher than Rinkeby ($0.0106$). This implies that Rinkeby is the most resilient to network partitioning (in terms of low graph modularity), and Ropsten is the least partitioning resilient.

Using the NetworkX tool~\cite{me:networkx:community}, we detect the communities of the Rinkeby and Goerli. 
In Rinkeby, there are four communities detected, and the biggest one (in green) are of $33.9\%$ of the nodes of the network. 
The average degree in the community is $52.3$, and $5.3\%$ of the nodes (i.e., 8 out of 151 nodes) only have a degree of 1.
In Goerli, there are seven communities detected, and the biggest one (in black) are of $24.6\%$ of the nodes of the network. 
The average degree in the community is $40.5$, and $2\%$ of the nodes (i.e., 5 out of 252 nodes) only have a degree of 1.

{\section{Discussion on the Impacts of EIP1559}
In EIP1559, there are three fee components: a base fee that is automatically set by the blockchain based on the recent block utilization, a priority fee set by the sender and a max fee also set by the sender. A transaction included in the blockchain always pays the base fee (burnt) and pays the priority fees to the miner. It is also ensured that the sum of the base fee and priority fee is lower than the max fee. 

Under EIP1559, the \texttt{mempool} uses the max fee to make admission/eviction decisions. Noteworthy is that when a pending transaction's max fee is below the base fee (i.e., negative priority fee), the transaction becomes underpriced and is dropped. Thus, in EIP1559, as long as we ensure the max fee in measurement transactions (i.e., $tx_A,tx_C,tx_O$) is above the base fee, the measurement process is not affected by the presence of EIP1559.
}